\documentclass[letterpaper]{JHEP3}

\usepackage{epsfig}
\usepackage{multicol}
\usepackage{wasysym}

\makeatletter
\def\gsim{\compoundrel>\over\sim}
\def\lsim{\compoundrel<\over\sim}
\def\compoundrel#1\over#2{\mathpalette\compoundreL{{#1}\over{#2}}}
\def\compoundreL#1#2{\compoundREL#1#2}
\def\compoundREL#1#2\over#3{\mathrel
         {\vcenter{\hbox{$\m@th\buildrel{#1#2}\over{#1#3}$}}}}
\makeatother

\newcommand{\be}{\begin{equation}}
\newcommand{\ee}{\end{equation}}
\newcommand{\bea}{\begin{eqnarray}}
\newcommand{\eea}{\end{eqnarray}}

\newcommand\fverbdo{\egroup\medskip\noindent%
			\fbox{\unhbox\fverbbox}\ }
\newcommand\fverbit{\egroup\item[\fbox{\unhbox\fverbbox}]}
\newbox\fverbbox

\title{Search for muon signal from dark matter annihilations in
  the Sun with the Baksan Underground Scintillator Telescope for 24.12 years} 

\author{M.M. Boliev\\
        Institute for Nuclear Research of Russian Academy of Sciences,\\
Baksan Neutrino Observatory, Kabardino-Balkariya 400900, Russia\\
	E-mail: \email{boliev2005@yandex.ru}}
\author{S.V. Demidov\\
	Institute for Nuclear Research of Russian Academy of  Sciences,\\ 
  prospect 60-th October 7A, Moscow 117312, Russia.\\
	E-mail: \email{demidov@ms2.inr.ac.ru}}
\author{S.P. Mikheyev$\dagger$\\
	Institute for Nuclear Research of Russian Academy of  Sciences,\\ 
  prospect 60-th October 7A, Moscow 117312, Russia.\\
        E-mail: \email{mikheyev@pcbai10.inr.ruhep.ru}}
\author{O.V. Suvorova\\
	Institute for Nuclear Research of Russian Academy of Sciences,\\
prospect 60-th October 7A, Moscow 117312, Russia.\\
	E-mail: \email{suvorova@cpc.inr.ac.ru}}

\abstract{We present a new dataset analysis of 
the
neutrino experiment at
  the Baksan Underground Scintillator Telescope with muon energy
  threshold about 1 GeV for the longest exposure time toward the Sun. In
  search for a signal from self-annihilations of dark matter particles
  in the center of the Sun we use an 
updated
sample of upward
  through-going muons for 24.12 years of live time. 
No observable excess has been found in measured muons
    relative to expected background from neutrinos of atmospheric
    origin. We present an improved data analysis procedure and
  describe it in detail. We set 
  the 90\% C.L. new upper limits on expected neutrino and muon fluxes
  from dark matter annihilations in the Sun, on the corresponding
  annihilation rates and cross sections of their elastic scattering
  off proton.}  

\dedicated{$\dagger$Dedicated to the memory of Stas Mikheyev.\\}

\begin{document} 


\section{Introduction}\label{sec:level1}
Evidence for a huge amount of missing (that is dark) mass in the
Universe comes from observed gravitational effects in astrometry, as it
was first discovered by F.Zwicky~\cite{Zwicky:33}. The most impressive 
hints come from measurements of rotation curves of spiral galaxies and
gravitational lensing (see
Refs.~\cite{Jungman:96,LBerg:00,Munoz:04,Silk:05}
for review) including reconstructed view of the merging process of
the bullet cluster
~\cite{WeakLens:06,WeakLens:12} which
  indicates that what is interpreted as dark matter effects can not be
explained by some modification of the gravity law. Recent high
precision measurements of microwave background
anisotropy~\cite{WMAP:11,CMB} continue to indicate that the
total energy density contains only about $4.6\%$ of baryonic matter
while other parts come from non ordinary forms of gravitating
matter: dark energy, about $72\%$, and dark matter (DM), about
$23\%$. The best fit of the results of observational astronomy favours
the flat $\Lambda CDM$ model in cosmology which implies existence of
the collisionless nonrelativistic DM in an extension of the Standard
Model (SM) of particle physics. These facts 
generate numerous multi-wavelength searches for relic particles
through their possible interactions with ordinary matter by direct and 
indirect methods. 
Moreover, several recent results obtained in low background detectors
of nuclear recoil
reactions~\cite{DAMA:08,Savage:2008er,CDMS:09,CoGent:10,CRESST:11}   
and in cosmic ray experiments of $\gamma$, $e^-e^+$ and $p\bar{p}$ 
measurements (see Ref.~\cite{Profum:10} for review) 
indicate on existence of spectral peculiarities which have been interpreted by
many authors as an evidence on existence of dark matter.
Indirect search for dark matter through neutrino
channel is accessible within dataset analysis of neutrino telescopes
in their regular observations of local sources like the Sun where DM
could be gravitationally trapped and further accumulated for the  time
of solar system age. 

Up-to now there are no hints on excess of neutrino events in the
direction towards the Sun as compared with expected background of
atmospheric neutrinos at all neutrino telescopes. However, these
observations allow to set upper limits on properties of dark
matter particles, in particular, upper limits on their
annihilation rates in the Sun which under certain assumptions can be 
translated into upper limits on the cross sections of elastic
scattering of dark matter on nucleons. Given the Sun chemical
composition with approximately $73\%$ of hydrogen the spin-dependent
elastic cross section of dark matter particles on proton appears to
be one of the most sensitive quantities in these searches as it can be
seen from the latest results from the
Super-Kamiokande~\cite{Tanaka:2011uf}, the  IceCube~\cite{:2012ef}
and the
ANTARES~\cite{Adrian-Martinez:2013ayv}
collaborations. 

The Baksan Underground Scintillator
Telescope~\cite{Baksan:79a} (further referred to as the Baksan or the
BUST) has muon energy threshold around 1~GeV allowing for searches
for neutrino signal from annihilation of relatively light dark matter
in the Sun. Corresponding minimal mass which can be probed at the Baksan is
about 10~GeV. 
Here we present updated results of the Baksan experiment with the
statistics twice as compared to the previous
ones~\cite{Baksan:97} and with new analysis improved in several
ways. First of all, current knowledge of neutrino properties has been
fully taken into account. Besides, the previous Baksan
results~\cite{Baksan:97,Baksan:96,Baksan:99} 
were obtained  in the framework of a specific model - MSSM (Minimal
Supersymmetric Standard Model) with neutralino as a dark matter
candidate (see Ref.~\cite{PDG:12} for a review). Performing a scan
over its parameter space but fixing the mass of dark matter particle 
conservative upper limits on muon flux and the rate of dark matter  
annihilation in the Sun have been obtained. In the present work rather
then studying a specific theory we follow a present day approach which
is applicable to a wider class of models. Namely, one can make an
assumption about dominating annihilation channel for dark matter
particles in the Sun and then set the limit on the corresponding
annihilation rate. This approach allows us to compare the Baksan
limits with the results of other neutrino experiments and at the same
time it 
is to a certain extent model independent: having the limits on 
possible annihilation channels allows one to apply them to a specific 
model. In our analysis we consider annihilations of dark matter
particles into 
$b\bar{b}$, ${\tau^+\tau^-}$ and ${W^+W^-}$. We use our own C code to
simulate neutrino propagation from the point of production in the Sun
to the detector level and compare our results with those obtained
with the help of WimpSim package~\cite{wimpsim,Blennow:2007tw}.

The paper is organized as follows. In Sec.~\ref{sec:level2} main
parameters of the Baksan Underground Neutrino Telescope and levels of
neutrino events selection are presented. In Sec.~\ref{sec:level3} we
describe the angular analysis of the upward going muon dataset. 
In Sec.~\ref{sec:level5} we describe in details numerical simulations
of neutrino propagation and muon signal in the telescope. The results
and discussion are presented in Sec.~\ref{sec:level6}. Finally,
Sec.~\ref{sec:level7} contains our conclusions.

\section{Experiment, triggers and sample of upward through-going
  muons}\label{sec:level2} 
The Baksan Underground Scintillator Telescope is a well known aged
neutrino telescope located in one of two underground tunnels of
the Baksan Neutrino Observatory. 
The BUST has 4-pi geometry for detection of penetrating charged
particles. 
The duration of continuous
measurements covers 34 years since December of 1978. Main parameters
of the telescope have been presented in details in
Refs.~\cite{Baksan:79a,Baksan:79b}. Separation of arrival
directions between up and down hemispheres is made by time-of-flight
(TOF) 
method with time resolution 5 ns~\cite{Baksan:79c}. It was 
shown~\cite{Baksan:79c} that in 95$\%$ of events values of inverse
reconstructed particle velocity $1/\beta$ lie in 
the
range of
$0.7\div1.3$ for single downward going muons. Such interval but
with negative sign of velocity is used  to select upward going muons
generated by neutrino interactions in down hemisphere.  
 
The telescope is located at the altitude 1700 m above see level in the 
Baksan valley of the North Caucasus and at the depth 850 hg/cm$^2$
under the mountain Andyrchi where the flux of atmospheric downgoing 
muons is reduced by factor of about 5000, but it is still higher than
upgoing muons flux by six orders of magnitude. Trajectories of
penetrating particles are reconstructed using the positions of hit
tanks, which represent together a system of 3150 liquid scintillation
counters of 
standard type $(70~{\rm cm} \times 70~{\rm cm} \times 30~{\rm cm})$ in 
configuration of parallelepiped $(17~{\rm m} \times 17~{\rm m} \times
11~{\rm m})$. The counters entirely cover all its sides and two
horizontal planes inside at the distances 3.6~m and 7.2~m from the
bottom. The planes are separated from each other by concrete absorber
of radiation length $\approx 160~{\rm g}/{\rm cm}^2$. The configuration
provides about $1.5^{\circ}$ of  muon angular resolution for
reconstructed trajectories longer than 7 meters.

The angular resolution of the Baksan telescope depends on a
  ratio of geometrical sizes of whole telescope and individual counter 
  (tank), as well as on selection criteria, in particular, on length
  of the muon trajectory. 
 The detailed studies have 
been done previously by Monte Carlo simulations and experimental data sets analysis of single downward 
going muons, pairs of parallel muons in muon groups~\cite{YMA:89,YMAprp:89}, the Moon shadowing 
of the cosmic rays~\cite{Karp:01,Karp:02}, the neutrino local sources studies~\cite{Zakid:96}, 
observations of muons toward the moving sources as for the Cygnus-X3 \cite{YMA:89}.

The rate of the general trigger of events at the telescope is 17~Hz.
There are two special hardware triggers used to select upward going
muons~\cite{Baksan:79b}. They reduce initial rate of downward going
muons approximately by factor of $10^3$. Trigger I covers the zenith
angle range $95^{\circ} \div 180^{\circ}$ while trigger II selects
horizontal muons in the angular range $80^{\circ} \div 100^{\circ}$. The
hardware trigger efficiency of 99\% has been measured with the flux of
atmospheric muons~\cite{Baksan:79b,Baksan:79c}. These two triggers
select 0.1\% of the initial rate, leaving about 1800 events per day 
for further processing. In the year 2000 the telescope data
acquisition system was upgraded and allowed to simplify the trigger
system down to one general trigger with the rate of 17~Hz. All raw
information is then undergone further selection using off-line
reconstruction code.  

Selection criteria of neutrino events in the Baksan neutrino
experiment has not been changed since the first data 
analyses~\cite{Baksan:99,Baksan:79b,Baksan:81,Baksan:06}. 
An off-line program reconstructs events and selects those of 
them which satisfy two simple requirements: presence of only one
reconstructed track of penetrating particle  and negative value of
measured velocity $\beta$. In addition, all events with negative
values of  $\beta$ have been scanned by the eyes to check possible
misinterpretation and there were no observed events within range of
$-0.5 \leq \beta \leq 0$. Also it is required that each trajectory has
an enter point lower than exit point on a range not less than the tank
size. For trajectories crossing only two scintillator planes (sample
of trigger II) we have excluded tracks having azimuthal angles
$0^{\circ} \leq \phi \leq 180^{\circ}$ from the direction of minimum
shallow depth to reduce background from downward going atmospheric
muons scattered at large angles in the rock.

The data used for the present analysis have been collected from
December of 1978 till November of 2009. In total, during 211275 hours
of live time (l.t.) which corresponds to 24.12 years, 1700 events
surviving the cuts described above have been collected. 

Additional cuts have been applied in order to reject upward-going
trajectories that could be mimicked by downward-going atmospheric muon
interactions or multiple muons.  i)~The muon trajectory should have
entry and exit points. Thus, stopping muons and neutrino interactions
inside the detector are excluded. ii)~Muon range inside detector must be
larger than 500~g$/$cm$^2$. These criteria cut off particles with
energy below  $\approx 1$~GeV. iii)~In the sample of events which
passed trigger II, those with entry or exit points closer than 1.5 m to
the plane edge are excluded. This cut reduces background from
atmospheric muons interactions. iv)~Finally, the events with
  $1/\beta$ in the range of $-1.3 \div -0.7$ only have been accepted. 

In total for 24.12 years of live time 1255 events survive all
cuts and are used in our further analysis. In
Fig.~\ref{ris2_RateBPST-TowardSun09-20b} (left) we show the rate of
collected measured upward through-going muons for good run
  periods of each year during all years of observation.  

\section{The Sun survey during three decades}\label{sec:level3}
In search for neutrinos from the dark matter annihilations in the Sun,
we analyse distribution of measured events as a function of
$\cos{\Psi_{\mu-{\rm Sun}}}$ where $\Psi_{\mu-{\rm Sun}}$ is the angle
between upward incoming muons and the Sun position. In
Fig.~\ref{ris2_RateBPST-TowardSun09-20b} (right) the 
obtained cosine distribution is presented. The mean rate per bin is
shown by blue line. We estimate the background expected from
atmospheric neutrinos directly from real data using shifted (false)
Sun positions. Here we follow our previous
analysis~\cite{Baksan:97,Baksan:96} where it was shown that
this method is compatible with Monte Carlo (MC) simulations of the
detector response on modeled atmospheric neutrino interactions
in the surrounded rock and muon propagation up to the detector. The
acceptance for these muons was calculated with the same requirements
for hardware triggers and same set of cuts as for real data. 
The detector efficiency as a function of muon energy is
shown~\cite{Baksan:97} in 
Fig.~\ref{Ris_EffcyBUST}. 
Mean energy of simulated atmospheric neutrinos, which produce muons
with energy larger 1~GeV for passing through the Baksan
telescope, was about 50~GeV~\cite{Baksan:97}. From comparison of data
sample for  
21.15~years of l.t.~\cite{Baksan:06} and collected MC statistics
larger than real data taking by factor of 22 (i.e., in total 460
years) it was found that the ratio of observed total number of events
to expected one without neutrino oscillations is $0.87 \pm 0.03(stat.) 
\pm 0.05(syst.) \pm 0.15(theor.)$. The details have been presented in
Ref.~\cite{Baksan:06}. 

At the location of the Baksan telescope (43,16$^\circ$N and
42,41$^\circ$E) the Sun is seen in average about half of time per day
during a year in both hemispheres; in
Fig.~\ref{SunTrack-ToFalseSun_Dw} (left) we show
the integral rate of 1255 selected upward through going muons as a
function of zenith angle  of the Sun for the BUST measured time of
good data periods (red points with error bars)  
and expected one for the detector runtime (blue histogram). The later
is normalized to this number of events. In calculation of the
positions of the Sun we apply the Positional Astronomy
Library~\cite{slalib}.  Performing the chi-square 
goodness-of-fit test we obtain a value of $\chi^2/d.o.f. =  10.38/19$.  
It means our multi-years measurements reproduce the Sun full-year
passing track with a good accuracy. That is the base for our further
analysis.

Potentially, the neutrinos from dark matter annihilations in the Sun
can be found at night time when the Sun is below horizon. We compare
zenith distributions of neutrino events coming at nightly ($N$) and
daily ($D$) time. Corresponding day-to-night asymmetry,
$\frac{N-D}{N+D}$ is shown in Fig.~\ref{SunTrack-ToFalseSun_Dw} (left)
and no significant difference between these data samples has been
observed. Angular distribution for ``night'' neutrinos is presented in
Fig.~\ref{AngLimToSun30-tot2009} (left). Note, that our sample
contains only upgoing events and the number of events in the bin
corresponding to the direction of the Sun $\cos{\Psi_{\mu-Sun}}=1$
remains the same is in Figure~\ref{ris2_RateBPST-TowardSun09-20b}
(right).  For "night" neutrinos, we compare cosine distribution in
Fig.~\ref{AngLimToSun30-tot2009} (left) and integrated angular 
distribution in Fig.~\ref{AngLimToSun30-tot2009} (right) with measured
background which has been obtained from averaged distribution of six
cases of false Sun positions shifted along the ecliptic. In
Fig.~\ref{AngLimToSun30-tot2009}~(right), the value of cone half-angle
$\gamma$ toward the Sun is shown on abscissa. We observe no excess of
events coming from the Sun. Assuming Poisson statistics for both
expected background $N_B$ and observed events $N_{obs}$ we obtain 90\%
C.L. upper limits on additional number of signal events $N_{S}^{90}$
as follows~\cite{PDG}: 
\begin{eqnarray}
\label{CL}
{C.L.} = 1 - \frac { e^{-(N_B+N^{90}_S)} 
\sum^{N_{obs}}_0{{(N_B+N^{90}_S)^n}\over{n!}} } {e^{-N_B}
\sum^{N_{obs}}_0{{{N_B}^n}\over{n!}}} 
\label{eq:five}.
\end{eqnarray}
Corresponding upper limits $N_{S}^{90}$ for each integer value of cone
half-angles in interval $1\div 25^{\circ}$ are also shown in 
Fig.~\ref{AngLimToSun30-tot2009}~(right).
The choice of the cone half-angle depends on the mass of dark
  matter particle and on particular annihilation channel. We discuss 
  it in Section~\ref{sec:level6}.

\section{Transport of neutrinos produced from dark matter
  annihilations in the Sun to the Earth}\label{sec:level5} 

In this Section we describe numerical procedure which we use to
simulate muon signal in the BUST resulting from dark matter annihilations
in the Sun. For this purpose the well known WimpSim package is
available~\cite{wimpsim,Blennow:2007tw}. However, to have more
flexibility we use our own C code for MC simulation of neutrino
propagation from the center of the Sun to the level of the Baksan
detector. Our general scheme of MC simulations is very similar to that
of WimpSim which is very well described in
Ref.~\cite{Blennow:2007tw}. They only differ in several details to be
discussed in a moment. 

In general, the initial spectrum of (anti)neutrino from dark matter
annihilations in the Sun is a mixture of inclusive (anti)neutrino
spectra from different annihilation channels
$\frac{dN_{\nu_{j}}}{dE_{\nu}} =
\sum_{i}B_{i}\frac{dN_{\nu_{j}}^{i}}{dE_{\nu}}$, where 
$\frac{dN_{\nu_{j}}^{i}}{dE_{\nu}}$ is the energy spectrum of $j$-th
type of  (anti)neutrino produced in $i$-th type of annihilation
channel. The quantities $B_{i}$ are the branching ratios of different
channels which can be calculated given a model for dark matter
candidate. Below we simulate neutrino propagation for several values
of masses of dark matter particle from 10~GeV to 1~TeV for three
dominant annihilation channels: $b\bar{b}$, $\tau^{-}\tau^{+}$ or
$W^{+}W^{-}$. The channel $b\bar{b}$ represents an example of ``soft''
spectrum while $W^{+}W^{-}$ and $\tau^{+}\tau^{-}$ are examples of
``hard'' spectra.

We do not calculate the neutrino spectra at the production point in
the Sun for each annihilation channel by ourselves but take them from
Ref.~\cite{Cirelli:2005gh}. We check that they coincide with a good
accuracy with those obtained with the help of WimpSim package except
for small deviations in spectra for $\nu_{\tau}$ and
$\bar{\nu}_{\tau}$. During the propagation of neutrinos from the
center of the Sun to the level of the detector we take into account
three-flavor neutrino oscillations in the vacuum, the Sun and Earth
including matter effects, absorption of neutrinos due to charged
current (CC) interactions with the interior of the Sun and Earth and
the change of the energy spectrum due to neutrino neutral current (NC)
elastic scattering and $\tau$-regeneration.
 
Neutrino oscillations in $3\times 3$ scheme are implemented according
to the algorithm described in
Refs.~\cite{Ohlsson:1999xb,Ohlsson:2001et}. According to this scheme
evolution operator analytically calculated for the case of constant
medium density is applied to the neutrino wave function. For the case
of varying density of the Sun one can subdivide the neutrino path into
sufficiently small intervals. Considering the medium density inside
each interval as a constant the total evolution operator can be
approximated as product of the evolution operators corresponding to
the sequence of intervals. We use solar model of
Ref.~\cite{Bahcall:2004pz}. For parameters of neutrino mixing matrix
we use the following values  
\begin{eqnarray}
\Delta m_{21}^2 = 7.62\cdot 10^{-5}~{\rm eV}^2,\;\;\;
|\Delta m_{31}^{2}| = 2.55\cdot 10^{-3}~{\rm eV}^2,\;\;\;
\\
\delta_{CP} = 0,\;\;\;
 \sin^2{\theta_{12}} = 0.32,\;\;\;
\sin^2{\theta_{23}} = 0.49,\;\;\;
\sin^2{\theta_{13}} = 0.026, \nonumber
\end{eqnarray}
which lie within $2\sigma $ range of experimentally allowed values
assuming normal hierarchy for neutrino masses~\cite{Tortola:2012te}. 

Our code follows straight trajectories of individual neutrinos
produced at some point near the center of the Sun which is simulated
according to the distribution~\cite{Cirelli:2005gh,Griest:1986yu} 
\be
n(r) = n_0{\rm exp}(-r^2/R_{\rm DM}^{2}),\; {\rm where}\;\; R_{\rm
  DM} = 0.01 R_{\astrosun}\sqrt{100~{\rm GeV}/m_{\rm DM}}.
\ee
To propagate neutrinos from the production point we calculate the
interaction length as 
\begin{equation}
L_{int}(E_{\nu}) =
\frac{1}{n_{n}\left(\sigma^{CC}_{n}(E_{\nu})+\sigma^{NC}_{n}(E_{\nu})\right)  
+ n_{p}\left(\sigma_{p}^{CC}(E_{\nu}) + \sigma_{p}^{NC}(E_{\nu})\right)},
\end{equation}
which depends on neutron and proton number densities in the Sun
$n_{n}$ and $n_{p}$, respectively, and the corresponding CC and NC
neutrino cross sections on nucleons $\sigma_{n,p}^{CC}$ and
$\sigma_{n,p}^{NC}$. Neutrino-nucleon DIS total cross sections and
corresponding energy and angular distributions are calculated
according to the expressions presented in Ref.~\cite{Paschos:2001np}
using CTEQ6~\cite{Pumplin:2002vw} parton distribution
functions. Inside the Sun, we choose the step size $\delta r$ as the
smallest between $R_{\astrosun}/F$ and $L_{int}/F$ where $F=300$ for 
$E_{\nu}\ge 100$~GeV and 
$F=3000$ for $E_{\nu}<100$~GeV. Smaller steps at lower energies are
required 
to take into consideration smaller oscillation length. We check that
total 
accuracy in neutrino fluxes obtained using these step sizes is better
then 1\%. Then, we calculate the 
probability for neutrino to have an interaction during this small step
$\delta r$: $P=1-{\rm exp}(-\delta r/L_{int})$. After that we simulate
the interaction process and if it takes place we determine its type
(CC or NC, proton or neutron) with the probabilities according to
corresponding cross sections and proton/neutron densities.  
When calculating CC neutrino-nucleon cross 
sections we include nonzero value of $\tau$-lepton mass while
electron and muon are considered as massless. In NC scattering process
the flavor structure of neutrino wave function is left
unchanged. Neutrino energy after this interaction is simulated
according to the corresponding differential cross section. For CC
interactions the flavor of neutrino  also has to be simulated and in
the case of $\nu_e$ and $\nu_{\mu}$ neutrinos we just drop them from
the flow. For $\nu_{\tau}$ we simulate the energy of resulting
$\tau$-lepton. Then the secondary neutrinos resulting from
$\tau$-decay are considered following procedure described in Appendix
of Ref.~\cite{Barger:2007xf} which is somewhat different from  WimpSim
implementation of this process\footnote{Secondary neutrinos contribute
  only a small part of   total neutrino flux and the differences
  between the two approaches   are not meaningfull. We thank Joakim
  Edsj\"o for correspondence on this point. 
}.

We check results obtained by our code for neutrino differential
fluxes at the level of the Earth (1 a.u.) with that of obtained by
WimpSim for the same spectra of neutrino at production in the Sun
calculated using WimpSim. The comparison is presented
in Figs.~\ref{wimpsim_comparison1},~\ref{wimpsim_comparison2}
and~\ref{wimpsim_comparison3} for three different annihilation
channels and different masses of dark matter particle. The results
agree with each other sufficiently well.  

High energy muons to be detected in neutrino telescope are produced in
CC interactions of neutrinos in the rock just below the detector. The
muons lose their energies during their passage from the production
point to the detector. The average muon energy loss is usually
parametrized by the following formula 
\be
\label{muon_en}
\langle \frac{dE}{dz} \rangle = -(\alpha(E) + \beta(E)E)\rho,
\ee
where $\rho$ is the density of the rock and the quantity $\alpha(E)$
corresponds to ionization energy loss while $\beta(E)$ accounts for
bremstrahlung, photonuclear interactions and pair production.
 We note that stochastic energy losses could be important for
  masses of dark matter particles $m_{DM}\gsim 1$~TeV (see,
  e.g.~\cite{Lipari:1991ut, Bugaev:1998bi}), however, we do
  not consider them in our calculations. 
We take into account the energy dependence of $\alpha(E)$ and
$\beta(E)$ to simulate the muon propagation in the rock. In the
interval $1\div 1000$~GeV of muon energies these coefficients
change~\cite{muons} from  $1.8\cdot10^{-3}$ to $2.7\cdot
10^{-3}$~GeV$\cdot$cm$^2/$g for $\alpha(E)$ and 
from $0.8\cdot 10^{-6}$ to $3.9\cdot 10^{-6}$~cm$^2/$g for
$\beta(E)$. 
Having rather moderate energy dependence in $\alpha(E)$
and $\beta(E)$ in many calculations constant values for these
quantities are used, for instance the values $\alpha(E)\approx
2.2\cdot  10^{-3}$~GeV$\cdot$cm$^2/$g  and $\beta(E)\approx 4.4\cdot 
10^{-6}$~cm$^2/$g for the rock are used in WimpSim. One could expect,
that for low energies $E_{\mu}\lsim 20$~GeV neglecting energy dependence 
for these coefficients leads to shorter muon ranges and some 
underestimate of muon flux. At the same time at high energies 
as $E_{\mu}\gsim 200\div 300$~GeV it should result in longer muon ranges and 
weak overestimate of muon flux.
In Fig.~\ref{muon_losses} we compare the both resulting muon energy spectra 
$d\Phi_{\mu}/dE_{\mu}$ calculated without and with energy dependence
in $\alpha(E)$ and $\beta(E)$ using values tabulated in Ref.~\cite{muons}.
We see that the obtained results confirm expectations discussed
above. Generally, we find that the correction due to the energy
dependence of $\alpha(E)$ and $\beta(E)$ can reach 18\% 
to the resulting muon flux in the rock from $b\bar{b}$ annihilation
channel for $m_{\rm DM}=10$~GeV and about 2-3\% to one from
  annihilation into $W^{+}W^{-}$ for $m_{\rm DM}=1000$~GeV. So, the
effect is more pronounced at low energies. In
Fig.~\ref{muon_ranges}~(left) we show the ratio of muon ranges
$R_{\mu}(E,E_{th})/R^{approx}_{\mu}(E,E_{th})$ where $R_{\mu}(E,E_{th})$ 
is calculated from Eq.~(\ref{muon_en}) taking into account energy
dependence in $\alpha(E)$ and  $\beta(E)$ while
$R^{approx}_{\mu}(E,E_{th})$ is obtained for constant values
$\alpha(E)\approx 2.2\cdot 10^{-3}$~GeV$\cdot$cm$^2/$g  and $\beta(E)\approx 
4.4\cdot 10^{-6}$~cm$^2/$g. 
As a cross-check we also verify our
simulations of muon flux by comparison with an analytical expressions
available for constant values of $\alpha(E)$ and $\beta(E)$ (see e.g. 
Ref.~\cite{Erkoca:2009by}).  
 In simulation of the muon flux from dark matter annihilation in
  the Sun we use the energy dependent coefficients $\alpha(E)$ and
  $\beta(E)$, taking into account slight widening of muon angle
  distribution due to multiple Coulomb
  scattering~\cite{PDG:12}.


\section{Results}\label{sec:level6} 

The upper limits on the number of signal events $N_{S}^{90}$ obtained from
the results of the Baksan Underground Scintillator Telescope (see
Section~3) can be used to put the 90\% upper limit on the flux 
$\Phi_{\mu}$ of muons with energies higher then 1~GeV from the dark
matter annihilation rate. This recalculation can be schematically
written as  
\be
\label{phi}
\Phi_{\mu}(90\%\; C.L.) = \frac{N_{S}^{90}}{\epsilon\times S_{eff}\times T}\;\;,
\ee
where $T$ is the live time, 
$\epsilon$ is a fraction of signal events inside corresponding cone
  half-angle
and $S_{eff}$ is effective area of Baksan telescope in the direction
of the expected signal which is averaged with Sun directions and also
includes the muon detection efficiency~\cite{Baksan:97,TB:91}.  
Corresponding effective exposure towards the Sun amounts to 
$0.49\times10^{15}~{\rm cm}^2\cdot{\rm s}$.  

The next step of the calculation is choice of a value of the cone
half-angle. Here we can explore obvious difference in angular
distribution of signal and background events. 
While muons from
atmospheric neutrinos are distributed nearly isotropic over the sky
, the
angular distribution of muons from dark matter annihilation is
expected to be correlated with the direction of the Sun. The angular
spread of the signal muons is determined by the hardness of 
neutrino spectra coming down to the Earth which in turn depends on the
DM mass and annihilation channel. Angular properties of the telescope
result in additional widening which we take into account with
Gaussian distribution over the direction of incoming muon with
incorporated angular resolution. The strategy used in the previous
analysis~\cite{Baksan:97} consisted of choosing the cone half-angle
which contains 90\% of events of upgoing muons expected from dark
matter annihilations in the Sun. However, one can try to optimize the
value of cone half-angle to reach the most stringent expected upper
limits. For this purpose we make use of model rejection factor (or
MRF) approach~\cite{Hill:2002nv}. We construct the quantity given by
the r.h.s of Eq.~(\ref{phi}) in which the upper limits on the number
of signal events $N_{S}^{90}$ are replaced by the upper limits
averaged over number of observed events using Poisson distribution,
$\bar{N}_{S}^{90}$. In this quantity $\bar{N}_{S}^{90}$, $\epsilon$
and $S_{eff}$ depend on the value of cone half-angle $\gamma$. Then,
we minimize it with respect to $\gamma$ and find the optimal value for
this angle. The obtained half-cone angles for three annihilation
channels $b\bar{b}$, $\tau^{+}\tau^{-}$  and $W^{+}W^{-}$ in
dependence on DM mass are shown in Fig.~\ref{muon_ranges}
(right). Found values of cone half-angles are used to calculate 
the upper limits at 90\% C.L. on the number of signal events
$N_{S}^{90}$ and muon flux from the Sun $\Phi_{\mu}$ using 
Eqs.~(\ref{CL}) and (\ref{phi}).


The muon flux is related to the dark matter annihilation rate
$\Gamma_{A}$ in the Sun as follows 
\be
\label{A_rate}
\Phi_{\mu} = \frac{\Gamma_{A}}{4\pi R^2}\times \sum_{\nu_{j}}
\int_{E_{th}}^{m_{\rm
    DM}}dE_{\nu_{j}}P(E_{\nu_{j}},E_{th})\frac{dN_{\nu_{j}}}{dE_{\nu_{j}}}, 
\ee
where $R=1$~a.u., $dN_{\nu_{j}}/dE_{\nu_{j}}$ is the spectrum of
$j$-th type neutrino at the Sun surface, $P(E_{\nu_{j}},E_{th})$ is
the probability to detect muon with energy $E_{\mu}>E_{th}=1$~GeV from 
$\nu_{j}$ neutrino with energy $E_{\nu_{j}}$ 
after its
  oscillations and propagation from the Sun surface while
the sum goes
over all types of neutrino and antineutrino. 
The probability $P(E_{\nu_{j}},E_{th})$  has been calculated 
  numerically
as described in the 
Section~4 for several masses of dark matter particles and annihilation
channels $b\bar{b}$, $\tau^{+}\tau^{-}$ and $W^{+}W^{-}$. 

In the case when the processes of dark matter capture and
annihilations in the Sun reach exact equilibrium during time
intervals much less than age of the solar system, 4.5 Giga years, the
annihilation rate $\Gamma_{A}$ should be equal to a half of capture
rate determined by scattering 
cross sections of dark matter particles off solar matter. In that
case, annihilation rate can be divided into two parts corresponding to
either spin-dependent (SD) or spin-independent (SI) type of
contribution to neutralino-nuclei interactions as follows   
\begin{eqnarray}
{\Gamma_{A}} = {\Gamma_{A}(\sigma_{SI})+\Gamma_{A}(\sigma_{SD})} 
\label{eq:four},
\end{eqnarray}
where $\Gamma_{A}(\sigma_{SI})$ and $\Gamma_{A}(\sigma_{SD})$ are
the parts of equilibrium annihilation rate determined by either
SI or SD interactions. Using the upper limits on the dark matter
annihilation rate in the Sun and hypothesis about exact 
equilibrium between capture and annihilation processes we can set
conservative upper limits on SD and SI elastic cross section of dark
matter particles on proton~\cite{Edsjo:09}. To reach this goal we use
recalculation procedure described in~\cite{Demidov:2010rq} where we
obtained corresponding coefficients in the following expressions 
\begin{eqnarray}
\sigma^{UppLim}_{SD}(m_{\chi})= \lambda^{SD}\left( m_{\chi}\right) \cdot
\Gamma^{UppLim}_A(m_{\chi})\\ 
\sigma^{UppLim}_{SI}(m_{\chi})= \lambda^{SI} \left( m_{\chi}\right) \cdot
\Gamma^{UppLim}_A(m_{\chi})
\end{eqnarray}
as functions of neutralino mass $m_{\chi}$. In obtaining
$\lambda^{SD}\left( m_{\chi}\right)$ and $\lambda^{SI} \left(
m_{\chi}\right)$ we assumed 0.3~GeV/cm$^{3}$ for the local density
of dark matter and 270 km/s for root-mean-square of the velocity
dispersion. For other details we refer reader to
Ref.~\cite{Demidov:2010rq}. We note that in~\cite{Demidov:2010rq}
  we used simplified solar model which allows to calculate the capture
  rate analytically~\cite{Gould:1992}. To be consistent with the
  present study we 
  recalculate the coefficients $\lambda^{SD}$ and $\lambda^{SI}$ for
  solar model~\cite{Bahcall:2004pz} which results in changes of these 
  coefficients within 10\% for SD cross section and within 15\% for
  SI. 
 
A possible influence of systematic effects on the number of observed
upward going muons have been tested during all years of the telescope
measurements. At the Baksan telescope there is implemented a
system of continuous online and offline diagnostics of each detector
elements and electronic channels, as well a regular recovering work at
the telescope planes one day in a week. The measurements have
shown \cite{Baksan:99} that the spread in threshold settings and the PMTs gain is
$\approx 10\%$ and the fraction of dead tanks is about $0.5\%$ over
the period of the observation. In a mean time the energy thresholds
of each of 3150 detectors have been tuned in an equal small magnitude
8 MeV,  while energy release of a relativistic particle is at least
six times larger. Thus possible macrofluctuations in any tank were not
seen by experimental monitoring of a single muon rate from upper
hemisphere~\cite{Karp:01,Karp:02}, while a tiny effect were observed due to enough
level of stability in the settings of the Baksan multiyears
measurements. It was shown ~\cite{BaksanVar:2013} the season variations in observed
rate of downgoing muons with magnitude value about $1\%$ over all
visible part of the sky. As known, these variations arise from changes
of stratosphere temperature. Instability of the detector parameters
during a long term operation of the telescope has been studied using
Monte Carlo simulations on the variation of PMTs gain ($\approx
10\%$), the spread in setting of thresholds ($\approx 10\%$), and time
off-set of PMTs ($\approx 2$~ns) as well as dead tanks ($\approx 1\%$)
\cite{Baksan:97,Baksan:99}. It was found that the number of detected events changes by
about $1\%$ given a variation of 10\% in the PMTs gain and
discrimination thresholds. 
In the measurements of upward going muons by TOF
method the time resolution of the telescope plays a more important
role. The time resolution is determined by intrinsic properties of the
scintillator  counters and PMTs and by the accuracy in adjusting the
time off-set of each tank. Before the telescope operation all counters
have been adjusted to an identical value of delay so that the maximal
spread of time off-sets appeared to be less than 2 ns. Continuous
monitoring of the time off-sets of all tanks and their online
performance shows that the spread increases and a tail at the level of
$\approx 2\%$ appears with time~\cite{Baksan:97, Baksan:99}. This type
of instability is 
considered as the main source of systematic uncertainties and is under
continuous studying. 
The systematic uncertainties of $8\%$ have
been evaluated by means of detector acceptance simulations, varying
parameters relevant to the observation of flux of upward going muons.  

Another sources of systematic uncertainties are related to estimation
of the background and simulation of the signal. Determination of the
background using shifted Sun positions results in uncertainty about
3\%. The extensive study of systematic uncertainties associated
to neutrino properties have been done. We perform MC simulation of
resulting muon flux using different oscillation parameters (within
$1\sigma$ range). In this study systematic uncertainties in the muon
flux are found to be of 8\% for $\tau^+\tau^-$ channel and of 5\% for
both $W^{+}W^{-}$  and $b\bar{b}$ channels regardless of neutrino
energy. Larger uncertainty for $\tau^+\tau^-$ indicates considerable
impact of neutrino oscillations in the case of this annihilation
channel (see also discussion of Fig.~\ref{osc} (right) further in the
Section). Further, we vary neutrino-nucleon cross sections using for
their uncertainties the results presented in
Ref.~\cite{CooperSarkar:2011pa} extrapolated for the case of low
energy neutrinos ($E_{\nu}<50$~GeV). Simulating resulting muon flux we
have found a spread of 10\% (4\%) for dark matter mass 10
(1000)~GeV. There is a considerable contribution from quasi-elastic 
neutrino interactions to neutrino-nucleon cross section in the low
energy range $\lsim 5$~GeV of neutrino which could be up to 50\% as it
was  shown, for example, in~\cite{Paschos:2001np} and which is
important for signal simulation in the case of the small DM masses 
about 10~GeV. We do not consider these contributions here; their
inclusion should increase total neutrino-nucleon cross section and
results in amplification of the signal and therefore in more stringent
upper limits. Thus, our limits for low masses are only conservative.  
Other uncertainties entering the recalculation to the upper limits on 
SD and SI cross sections are related to nuclear
form-factors, solar composition, influence of planets on capture of
dark matter particles, structure of the dark matter halo and the dark
matter velocity distribution and they are discussed in
Refs.~\cite{Edsjo:09,Rott:2011fh}. Also we note that recent studies
indicate a somewhat higher value of the local dark matter density,
closer to 0.4~GeV/cm$^3$~\cite{Catena:2009mf}. However, this increase
will make our limits only stronger.

Incorporation of systematic uncertainties into the final results on
the mentioned upper limits has been performed by the 
method of Cousins~\&~Highland~\cite{Cousins:1991qz}. The results for
upper limits on quantities discussed above are collected in
Table~\ref{table:lambda} for several masses of dark 
matter and for chosen annihilation channels. Here we present obtained
optimized values of the half-cone angle $\gamma$ and the upper
limits at 90\% C.L. on following quantities: the number of signal
events $N_{S}^{90}$, 
muon flux from the Sun $\Phi_{\mu}$ in the rock, annihilation rate in
the Sun $\Gamma_{A}$, the spin-dependent $\sigma^{SD}_{\chi p}$ and
spin-independent $\sigma^{SI}_{\chi p}$ scattering cross sections of
dark matter particle off proton and also upper limits on neutrino
fluxes $\Phi_{\nu}$ integrated from energy $E_{\nu}=1$~GeV. 

In Fig.~\ref{osc} (left) are shown the 90\% C.L. upper limits on muon
fluxes from the Baksan's data for the annihilation channels
$W^{+}W^{-}$, $b\bar{b}$ and $\tau^{+}\tau^{-}$ along with
corresponding limits of Super-Kamiokande~\cite{Tanaka:2011uf},
IceCube~\cite{:2012ef} and
ANTARES~\cite{Adrian-Martinez:2013ayv}.  
Typically, for indirect searches with neutrino telescopes the
annihilation channel $\tau^{+}\tau^{-}$ are not considered separately
from $W^{+}W^{-}$ to set limits on dark matter properties. However,
the difference between neutrino spectra from DM annihilation into
$\tau^{+}\tau^{-}$ and $W^{+}W^{-}$ plays an important role. In
spite of the fact that the limits on muon fluxes are approximately the
same for $W^{+}W^{-}$ and $\tau^{+}\tau^{-}$, the limits on SD cross
section (and on the corresponding annihilation rate)  are quite
different for these two channels as one can see in
Figs.~\ref{sd_lim}. The explanation of this fact resides partly in the
effect of neutrino oscillations: the enhancement due to this effect
for $\tau^{+}\tau^{-}$ channel is considerably larger than for the
case of $W^{+}W^{-}$ channel. This is illustrated in Fig.~\ref{osc}
(right) where the ratio of muon fluxes with and without oscillation is
shown for three annihilation channels depending on mass of dark
matter. Another important effect is a breaking "democracy" in average
number of high energy neutrinos per one act of dark matter
annihilations: for $\tau^{+}\tau^{-}$ pairs it is larger in a factor
of $2\div3$ than for $W^{+}W^{+}$ or $b\bar{b}$ channels. It can be
seen in Fig.~\ref{osc}(right) that neutrino oscillations result in a
decrease of muon flux from $b\bar{b}$ annihilation channel, which is
expected to be dominating at lighter neutralino masses. At the same
time the muon flux from $W^{+}W^{-}$ annihilations is enhanced.  
And finally, using $\tau^{+}\tau^{-}$ channel allows to extend the
upper limits on SD cross section for ``hard'' type of neutrino spectra
to lower masses of dark matter. 

The Baksan 90\% C.L. upper limits on SD cross section are shown in
Fig.~\ref{sd_lim} in comparison with experimental results from direct
and indirect DM searches. Here we show recent limits of
Super-Kamiokande~\cite{Tanaka:2011uf}, IceCube~\cite{:2012ef},
ANTARES~\cite{Adrian-Martinez:2013ayv,ANTARES:2012,Coyle:2012wv},
PICASSO~\cite{Archambault:2012pm}, KIMS~\cite{Kim:2012rza},
SIMPLE~\cite{Felizardo:2011uw}, CMS~\cite{Chatrchyan:2012me} and
ATLAS~\cite{ATLAS:2012ky} collaborations. A signal range of
DAMA~\cite{DAMA:08,Savage:2008er} experiment is also
depicted. Similarly, the Baksan 90\% C.L. upper limits on SI cross
section are presented in Fig.~\ref{si_lim} in comparison with results
of IceCube~\cite{:2012ef}, ANTARES~\cite{Adrian-Martinez:2013ayv},
DAMA~\cite{DAMA:08,Savage:2008er}, CoGeNT~\cite{CoGent:10},
XENON100~\cite{Aprile:2013doa} and CDMS~\cite{CDMS:09}. 
Note that the hard channel for the IceCube limits shown in these
  figures is assumed annihilation into $W^{+}W^{-}$ for $m_{DM}>m_{W}$
  and into   $\tau^{+}\tau^{-}$ otherwise.
As one can see 
from the Figures the Baksan has the strongest limits at low DM masses
for $\tau^{+}\tau^{-}$  annihilation channel as compared to the other
neutrino telescopes. 

Let us note, that the indirect  limits on SD and SI cross sections
are not fully model independent. The limits of neutrino
telescopes strongly rely on hypothesis about equilibrium between
capture and annihilation  processes with dark matter particles in the
Sun. However, the equilibrium condition is fulfilled in most of
the dark matter scenarios. As for the impressive collider bounds by
CMS~\cite{Chatrchyan:2012me} and ATLAS~\cite{ATLAS:2012ky}
collaborations on dark matter interactions they are obtained by a
recalculation from direct limits on cross sections of dark matter pair
production in association with di-jet and missing transverse
energy. There are also similar collider bounds from searches using
photon and missing transverse energy
signature~\cite{Aad:2012fw,Chatrchyan:2012tea}. This
recalculation~\cite{Goodman:2010ku,Fox:2011pm,Rajaraman:2011wf} is
based on the fact that the diagrams describing the process of dark
matter pair production in $pp$-collisions and elastic scattering 
process of DM off proton are related by crossing symmetry. However,
these procedure and corresponding indirect bounds on SI and SD
cross sections are very model dependent because of different
kinematics of these processes. The collider bounds especially weaken
in the models when dark matter interacts with quarks via light
mediators~\cite{Goodman:2010ku,Fox:2011pm,Rajaraman:2011wf} which is 
explained by completely different scaling of dark matter pair
production and dark matter -nucleon scattering cross sections with
c.o.m. energy and mass of mediator. Therefore, collider limits on SI
and SD elastic cross sections of dark matter particle off proton
should be interpreted with care.  Finally, we stress that given a
particular model  different kinds of experiments on DM searches give
complementary bounds on the properties of dark matter particles.

\TABLE[!ht]{
\footnotesize
\begin{tabular}{|ccccccccc|}
\hline\hline
$m_{\rm DM},${\rm GeV}$ $ & channel & $\gamma$, deg &
$N_s^{90}$ & $\Phi_{\mu}$,${\rm cm}^{-2} {\rm s}^{-1}$ &
$\Gamma_{A}$, ${\rm s}^{-1}$ & $\sigma^{SI}_{\chi p}$, pb & 
$\sigma^{SD}_{\chi p}$, pb  & $\Phi_{\nu_{\mu}}$,${\rm cm}^{-2} {\rm s}^{-1}$
\\\hline
10.0 & $b\bar{b}$ & 15.0 & 9.6 & $4.9\cdot 10^{-14}$ & $3.3\cdot
10^{26}$ & $1.2\cdot 10^{-3}$ & $7.0\cdot 10^{-2}$  & 0.021\\
     & $\tau^{+}\tau^{-}$ & 11.7 & 6.9 & $2.7\cdot 10^{-14}$ & $8.1\cdot
10^{24}$ & $2.9\cdot 10^{-5}$ & $1.7\cdot 10^{-3}$ & 0.0022  \\
\hline
30.0 & $b\bar{b}$ & 10.7 & 4.6 & $1.7\cdot 10^{-14}$ & $1.0\cdot
10^{25}$ & $6.2\cdot 10^{-5}$ & $1.0\cdot 10^{-2}$ & $9.4\cdot 10^{-4}$ \\
     & $\tau^{+}\tau^{-}$ & 7.4 & 4.1 & $1.3\cdot 10^{-14}$ & $4.1\cdot
10^{23}$ & $2.4\cdot 10^{-6}$ & $4.0\cdot 10^{-4}$ & $1.2\cdot 10^{-4}$ \\
\hline
50.0 & $b\bar{b}$ & 8.8 & 4.2 & $1.4\cdot 10^{-14}$ & $3.6\cdot
10^{24}$ & $3.2\cdot 10^{-5}$ & $8.2\cdot 10^{-3}$ & $3.4\cdot 10^{-4}$ \\
     & $\tau^{+}\tau^{-}$ & 6.4 & 4.4 & $1.4\cdot 10^{-14}$ & $1.6\cdot
10^{23}$ & $1.4\cdot 10^{-6}$ & $3.6\cdot 10^{-4}$ & $4.7\cdot 10^{-5}$ \\
\hline
70.0 & $b\bar{b}$ & 7.8 & 4.6 & $1.5\cdot 10^{-14}$ & $2.0\cdot
10^{24}$ & $2.6\cdot 10^{-5}$ & $8.5\cdot 10^{-3}$ & $2.1\cdot 10^{-4}$ \\
     & $\tau^{+}\tau^{-}$ & 5.8 & 4.7 & $1.4\cdot 10^{-14}$ & $8.6\cdot
10^{22}$ & $1.1\cdot 10^{-6}$ & $3.6\cdot 10^{-4}$ & $2.6\cdot 10^{-5}$ \\
\hline
90.0 & $b\bar{b}$ & 7.3 & 4.0 & $1.3\cdot 10^{-14}$ & $1.2\cdot
10^{24}$ & $2.0\cdot 10^{-5}$ & $7.9\cdot 10^{-3}$ & $1.2\cdot 10^{-4}$ \\
     & $\tau^{+}\tau^{-}$ & 5.4 & 3.9 & $1.1\cdot 10^{-14}$ & $4.5\cdot
10^{22}$ & $7.6\cdot 10^{-7}$ & $3.0\cdot 10^{-4}$ & $1.3\cdot 10^{-5}$ \\
     & $W^{+}W^{-}$ & 5.5 & 3.8 & $1.1\cdot 10^{-14}$ & $1.0\cdot
10^{23}$ & $1.7\cdot 10^{-6}$ & $6.9\cdot 10^{-4}$ & $1.2\cdot 10^{-5}$ \\
\hline
100.0 & $b\bar{b}$ & 7.1 & 4.1 & $1.3\cdot 10^{-14}$ & $1.0\cdot
10^{24}$ & $1.9\cdot 10^{-5}$ & $8.4\cdot 10^{-3}$ & $1.0\cdot 10^{-4}$ \\
     & $\tau^{+}\tau^{-}$ & 5.3 & 3.9 & $1.1\cdot 10^{-14}$ & $3.8\cdot
10^{22}$ & $7.1\cdot 10^{-7}$ & $3.1\cdot 10^{-4}$ & $1.1\cdot 10^{-5}$ \\
     & $W^{+}W^{-}$ & 5.4 & 3.8 & $1.1\cdot 10^{-14}$ & $8.7\cdot
10^{22}$ & $1.6\cdot 10^{-6}$ & $7.1\cdot 10^{-4}$ & $9.4\cdot 10^{-6}$ \\
\hline
200.0 & $b\bar{b}$ & 5.9 & 4.6 & $1.4\cdot 10^{-14}$ & $3.9\cdot
10^{23}$ & $1.7\cdot 10^{-5}$ & $1.2\cdot 10^{-2}$ & $3.9\cdot 10^{-5}$ \\
     & $\tau^{+}\tau^{-}$ & 4.4 & 3.2 & $8.7\cdot 10^{-15}$ & $9.9\cdot
10^{21}$ & $4.4\cdot 10^{-7}$ & $3.0\cdot 10^{-4}$ & $2.7\cdot 10^{-6}$ \\
     & $W^{+}W^{-}$ & 4.3 & 3.2 & $8.5\cdot 10^{-15}$ & $2.4\cdot
10^{22}$ & $1.1\cdot 10^{-6}$ & $7.2\cdot 10^{-4}$ & $2.4\cdot 10^{-6}$ \\
\hline
300.0 & $b\bar{b}$ & 5.4 & 3.9 & $1.1\cdot 10^{-14}$ & $1.8\cdot
10^{23}$ & $1.5\cdot 10^{-5}$ & $1.2\cdot 10^{-2}$ & $2.0\cdot 10^{-5}$ \\
     & $\tau^{+}\tau^{-}$ & 4.1 & 3.6 & $8.3\cdot 10^{-15}$ & $5.4\cdot
10^{21}$ & $4.7\cdot 10^{-7}$ & $3.2\cdot 10^{-4}$ & $1.5\cdot 10^{-6}$ \\
     & $W^{+}W^{-}$ & 4.0 & 3.3 & $8.4\cdot 10^{-15}$ & $1.4\cdot
10^{22}$ & $1.1\cdot 10^{-6}$ & $9.5\cdot 10^{-4}$ & $1.4\cdot 10^{-6}$ \\
\hline
500.0 & $b\bar{b}$ & 4.9 & 4.0 & $1.1\cdot 10^{-14}$ & $9.8\cdot
10^{22}$ & $1.7\cdot 10^{-5}$ & $1.8\cdot 10^{-2}$ & $1.0\cdot 10^{-5}$\\
     & $\tau^{+}\tau^{-}$ & 3.9 & 3.3 & $8.6\cdot 10^{-15}$ & $5.6\cdot
10^{21}$ & $5.6\cdot 10^{-7}$ & $5.9\cdot 10^{-4}$ & $8.4\cdot 10^{-7}$ \\
     & $W^{+}W^{-}$ & 3.5 & 3.3 & $8.3\cdot 10^{-15}$ & $9.3\cdot
10^{21}$ & $1.6\cdot 10^{-6}$ & $1.7\cdot 10^{-3}$& $8.4\cdot 10^{-7}$  \\
\hline
700.0 & $b\bar{b}$ & 4.6 & 3.1 & $8.8\cdot 10^{-15}$ & $5.9\cdot
10^{22}$ & $1.7\cdot 10^{-5}$ & $1.8\cdot 10^{-2}$ & $5.9\cdot 10^{-6}$ \\
     & $\tau^{+}\tau^{-}$ & 3.8 & 3.3 & $8.3\cdot 10^{-15}$ & $2.5\cdot
10^{21}$ & $7.6\cdot 10^{-7}$ & $8.8\cdot 10^{-4}$ & $6.4\cdot 10^{-7}$ \\
     & $W^{+}W^{-}$ & 3.7 & 3.3 & $8.7\cdot 10^{-15}$ & $7.9\cdot
10^{21}$ & $2.5\cdot 10^{-6}$ & $2.9\cdot 10^{-3}$ & $6.7\cdot 10^{-7}$ \\
\hline
900.0 & $b\bar{b}$ & 4.5 & 3.1 & $8.8\cdot 10^{-15}$ & $6.2\cdot
10^{22}$ & $3.0\cdot 10^{-5}$ & $3.7\cdot 10^{-2}$ & $6.1\cdot 10^{-6}$ \\
     & $\tau^{+}\tau^{-}$ & 3.7 & 3.3 & $8.3\cdot 10^{-15}$ & $2.1\cdot
10^{21}$ & $1.0\cdot 10^{-6}$ & $1.3\cdot 10^{-3}$ & $5.5\cdot 10^{-7}$ \\
     & $W^{+}W^{-}$ & 3.6 & 3.3 & $8.4\cdot 10^{-15}$ & $7.5\cdot
10^{21}$ & $4.3\cdot 10^{-6}$ & $5.4\cdot 10^{-3}$ & $7.5\cdot 10^{-7}$ \\
\hline
1000.0 & $b\bar{b}$ & 4.5 & 3.1 & $8.8\cdot 10^{-15}$ & $6.2\cdot
10^{22}$ & $3.4\cdot 10^{-5}$ & $4.2\cdot 10^{-2}$ & $5.6\cdot 10^{-6}$ \\
     & $\tau^{+}\tau^{-}$ & 3.6 & 3.4 & $8.4\cdot 10^{-15}$ & $2.1\cdot
10^{21}$ & $1.2\cdot 10^{-6}$ & $1.5\cdot 10^{-3}$ & $5.4\cdot 10^{-7}$ \\
     & $W^{+}W^{-}$ & 3.4 & 3.4 & $8.4\cdot 10^{-15}$ & $7.4\cdot
10^{21}$ & $4.3\cdot 10^{-6}$ & $5.4\cdot 10^{-3}$ & $5.8\cdot 10^{-7}$ \\
\hline\hline
\end{tabular}
\caption{\label{table:lambda}
 Half-cone angles $\gamma$, upper 90\% limits on the number of signal
 events $N_{S}^{90}$,  the muon flux $\Phi_{\mu}$, the dark matter
 annihilation rate in the Sun $\Gamma_{A}$, the dark matter -proton
 spin-dependent  $\sigma^{SD}_{\chi p}$ and spin-independent
 $\sigma^{SI}_{\chi p}$ scattering cross sections and neutrino
 fluxes $\Phi_{\nu}$.} 
}

\section{Summary}\label{sec:level7}

We have performed updated analysis of indirect dark matter search
with the Baksan Underground Scintillator Telescope data for 24.12
years of live time. In search for an excess of upward going muons in the
direction toward the Sun we do not observe any significant deviation
from expected atmospheric background. We have presented the 90\%
C.L. upper limits on spin-dependent and spin-independent elastic cross
section of dark matter on proton assuming particular annihilation
channels $b\bar{b}$, $W^{+}W^{-}$ and $\tau^{+}\tau^{-}$. These limits
have been derived from the limits on muon fluxes and annihilation
rates for dark matter masses in the interval $10\div1000$~GeV. The
best value for the upper limit on SD elastic cross section is about
$3\times 10^{-4}$~picobarn for masses of dark matter particle within
$100\div 200$~GeV, that is comparable 
with those presented by other operating neutrino telescopes.
At the same time in the range of low masses of dark matter particle 
$10\div 50$~GeV the Baksan limits on SD elastic cross section in the
case of $\tau^{+}\tau^{-}$ annihilation channel are the most stringent
to date among the results obtained by neutrino telescopes. 

\begin{acknowledgments}
We acknowledge our colleagues from the Baksan Observatory V.Petkov,
Yu.Novoseltsev, R.Novoseltseva, V.Volchenko and A.Yanin
for partnership in providing a long-term stability of the operating
telescope. S.D. thanks D.Gorbunov for valuable discussions. 
We are grateful to a referee for very valuable advices and comments. 
We also acknowledge support from Federal Program "Researches and
developments  of priority directions of science and technology in
Russia" under contract No.~16.518.11.7072. The work of S.D. was
supported in part by the grants of the President of the Russian 
Federation NS-5590.2012.2, MK-2757.2012.2, by Russian Foundation for
Basic Research grants 11-02-01528-a, 12-02-31726-mol-a and
13-02-01127Á and by the 
Ministry of Science and Education under contract No.~8412. The
numerical part of the work was performed on Calculational Cluster of
the Theory Division of INR RAS. 
\end{acknowledgments}

%
\FIGURE[htb]{
\begin{tabular}{cc}
\includegraphics[angle=0,width=0.50\columnwidth]{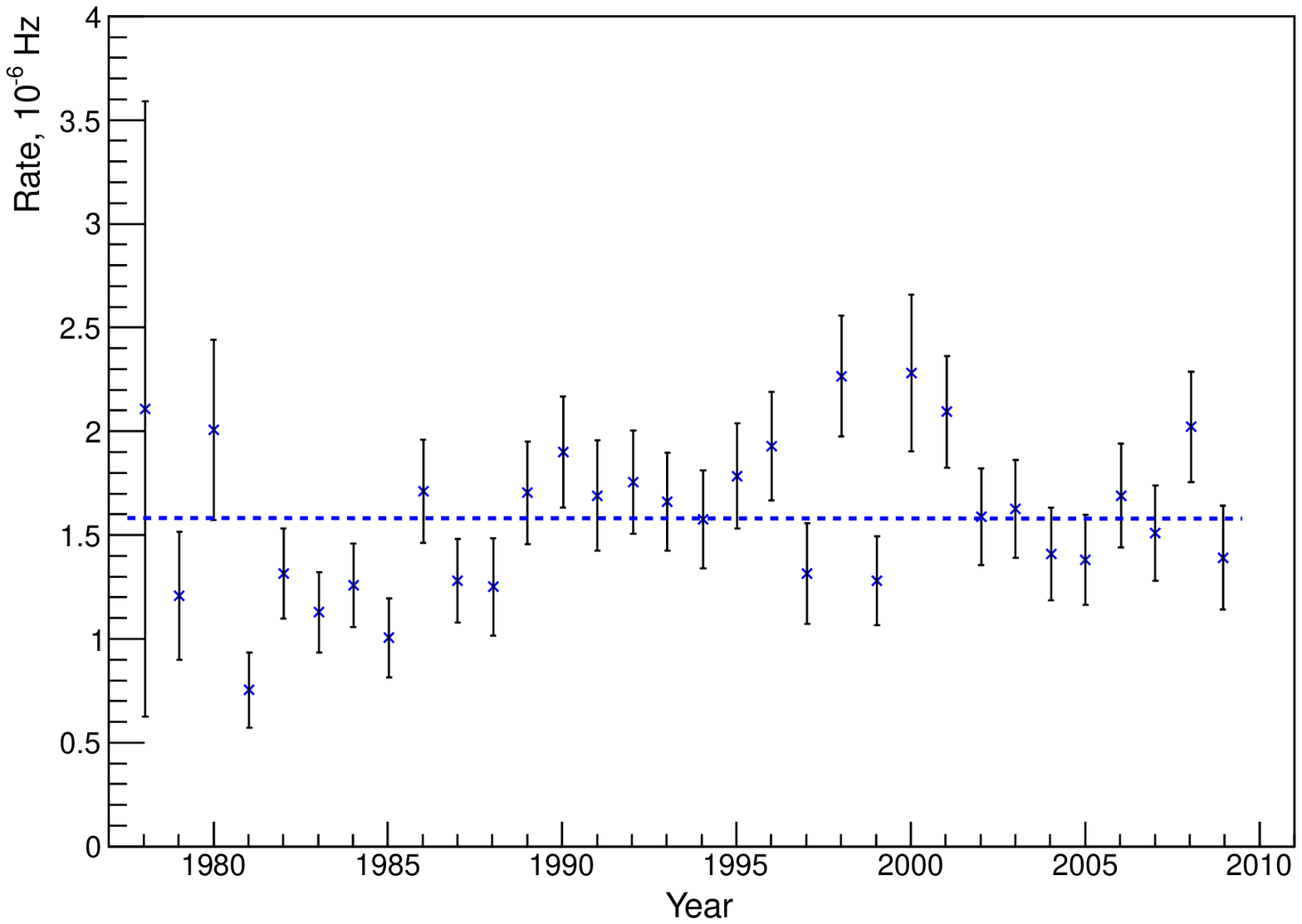}
\includegraphics[angle=0,width=0.50\columnwidth]{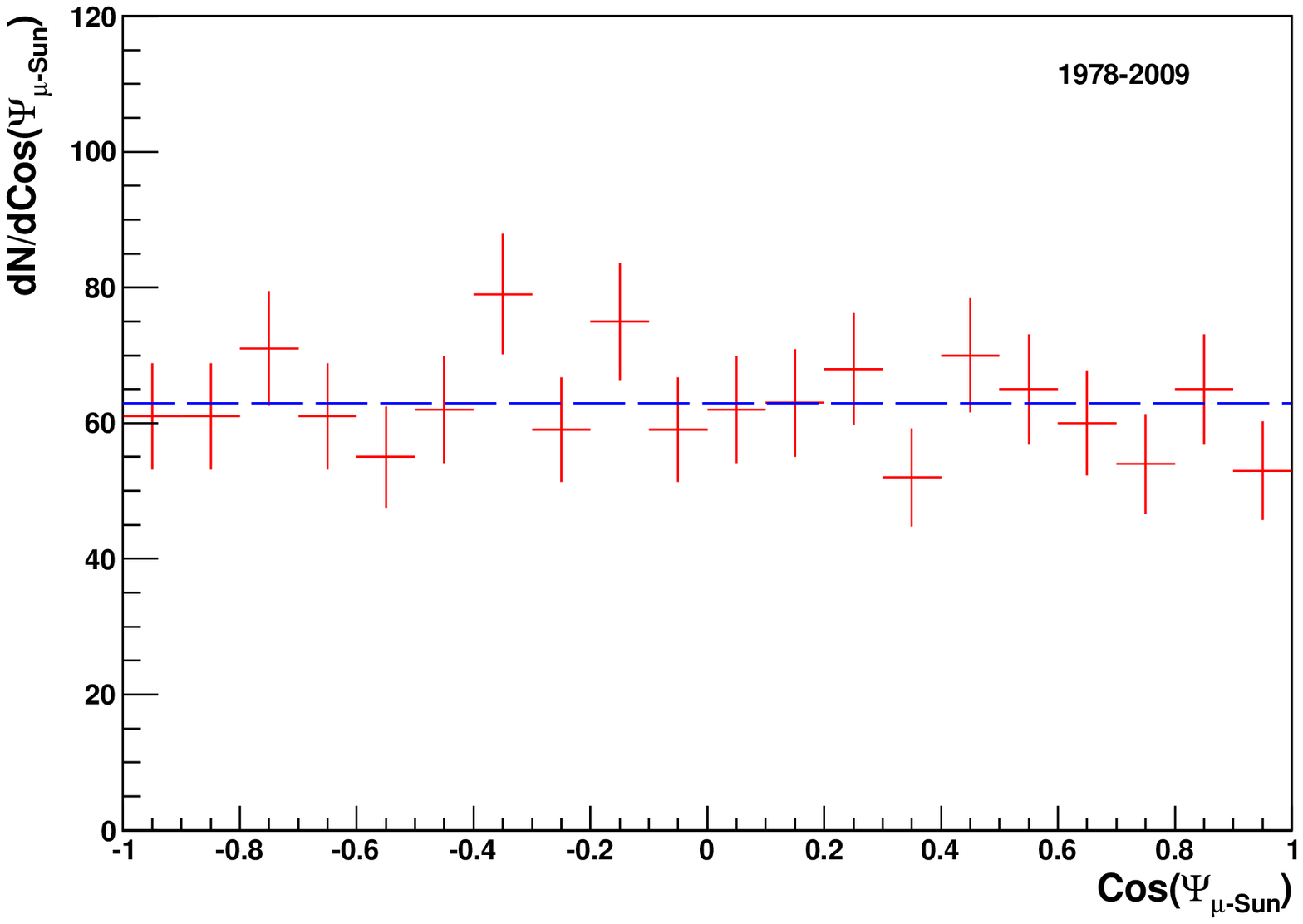}
\end{tabular}
\caption{\label{ris2_RateBPST-TowardSun09-20b} Left: Measured rate of
 upward through going muons at the BUST since December 1978. Right:
 Distribution of measured events in angle $\Psi_{\mu{\rm -Sun}}$
 between incoming muon events at the BUST and the Sun position. The 
direction of the Sun corresponds to $\cos{\Psi_{\mu{\rm -Sun}}}=1$.}  
}
%
\FIGURE[htb]{
\includegraphics[angle=0,width=0.7\columnwidth]{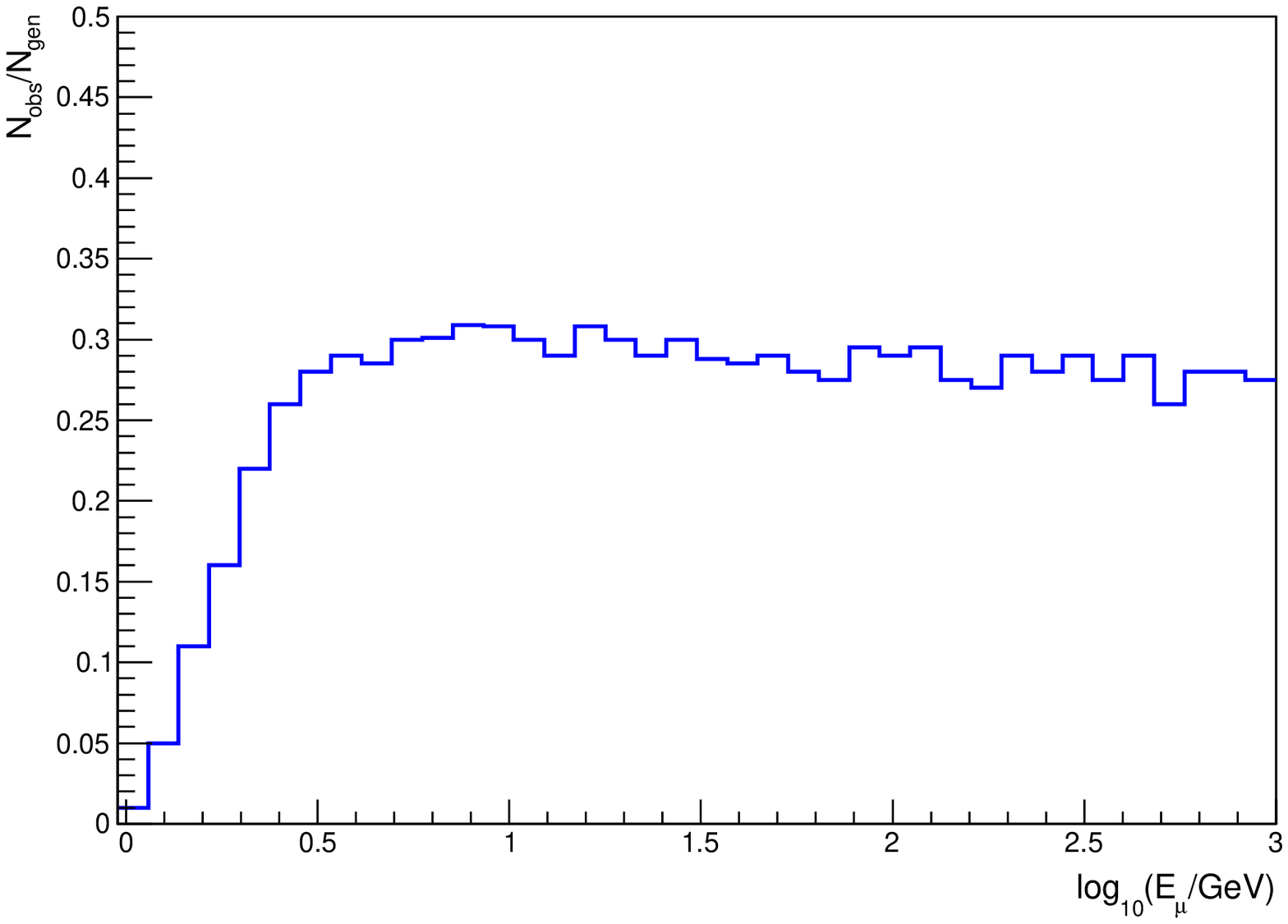}
\caption{\label{Ris_EffcyBUST} 
The ratio of the number of MC events which survive all cuts
($N_{obs}$) and total MC generated neutrino interactions ($N_{gen}$)
in the Baksan surrounding rock in dependence on muon energy.}  
}
\FIGURE[htb]{
\begin{tabular}{cc}
\includegraphics[angle=0,width=0.50\columnwidth]{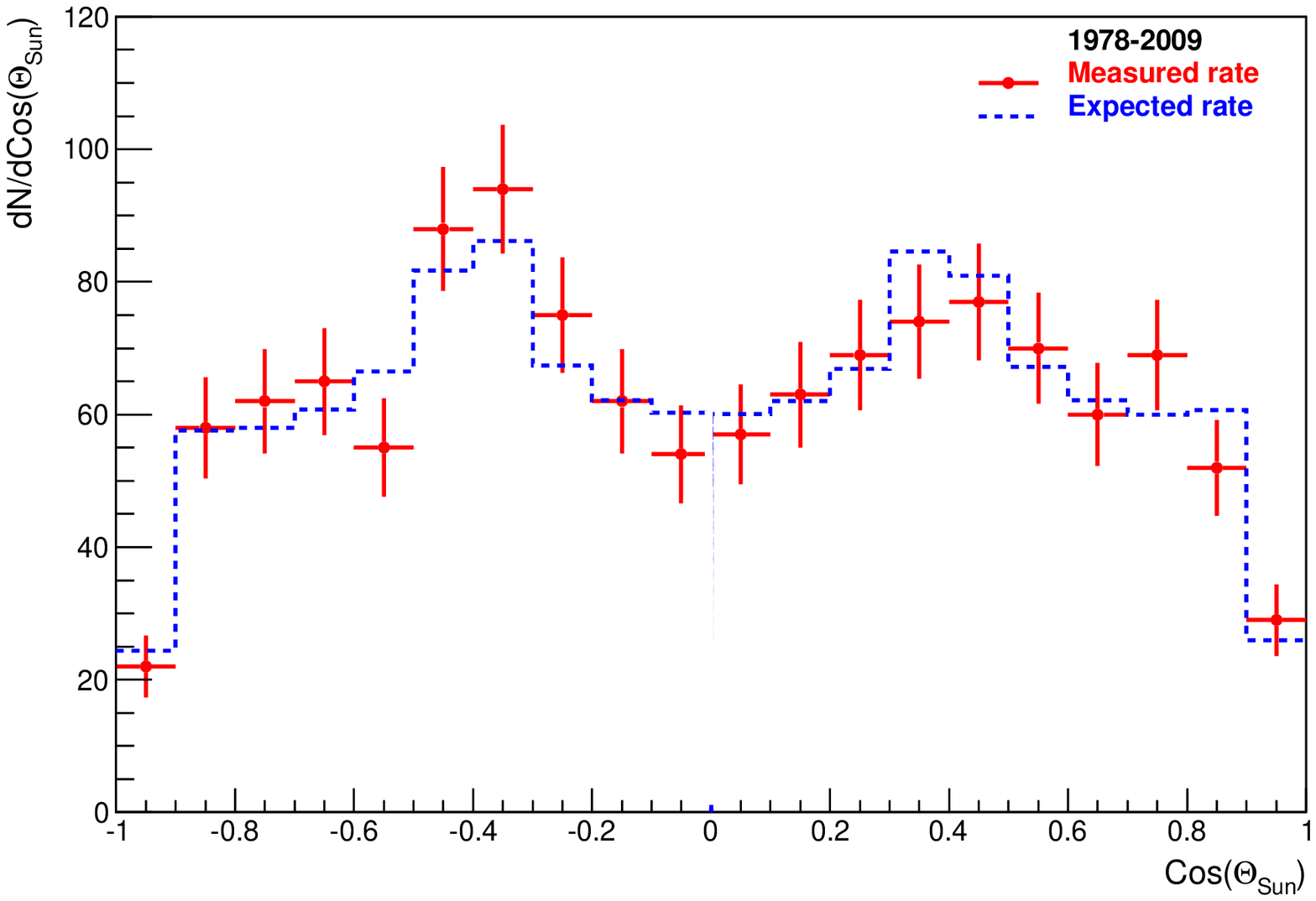} 
&
\includegraphics[angle=0,width=0.50\columnwidth]{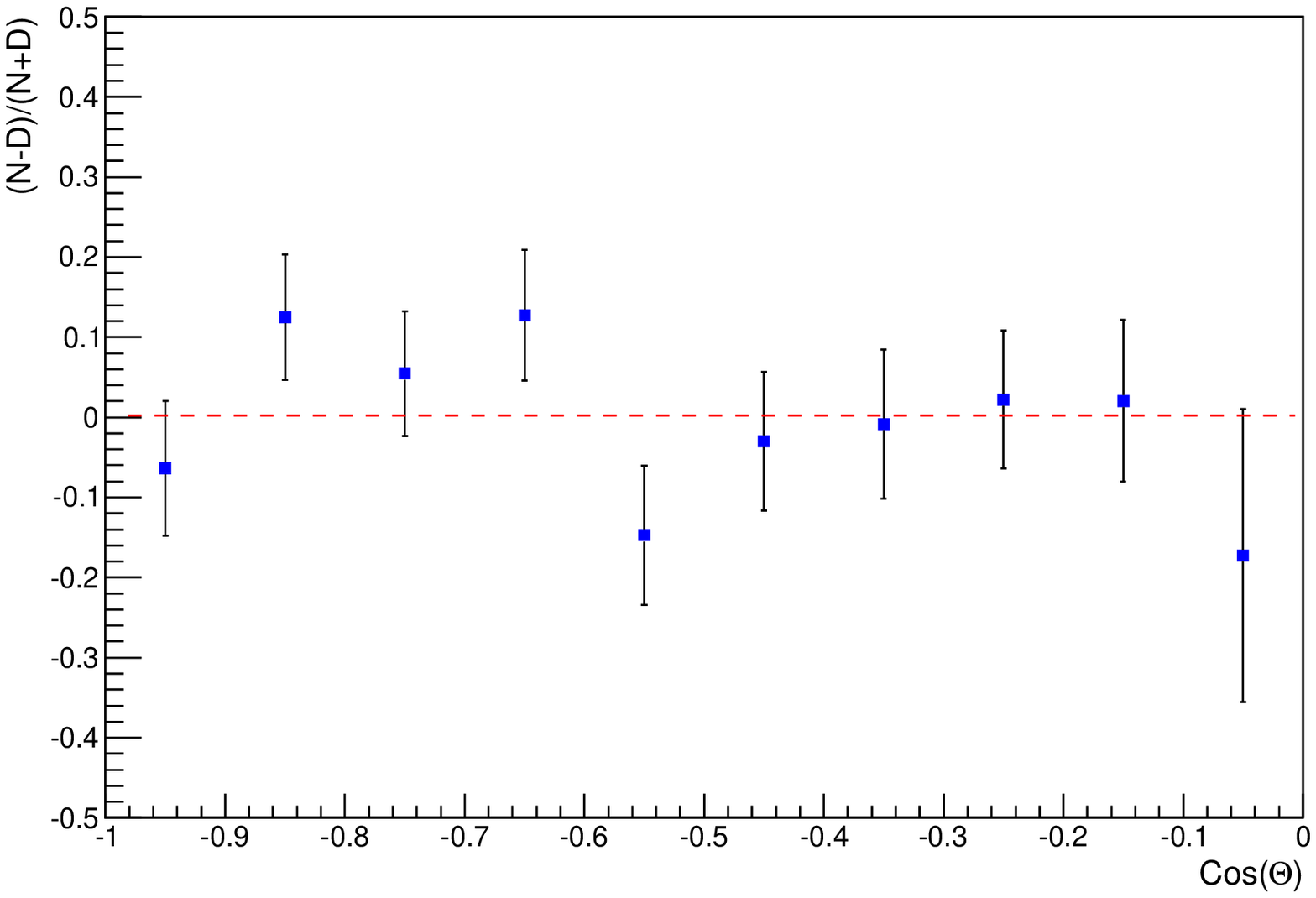} 
\end{tabular}
\caption{\label{SunTrack-ToFalseSun_Dw} Left: 
  The integral rate of 1255 selected upward through going muons as
  a function of zenith angle of the Sun for the BUST measured time of
  good data periods (red points with error bars) and expected one for
  the detector runtime (blue histogram). The later is normalized to
  this number of events.  
 Right: Zenith distribution for asymmetry $\frac{N-D}{N+D}$ between the number
of neutrino events coming at time when the Sun is under (night, $N$)
or above (day, $D$) the horizon.
}}
\FIGURE[htb]{
\begin{tabular}{cc}
\includegraphics[angle=0,width=0.47\columnwidth]{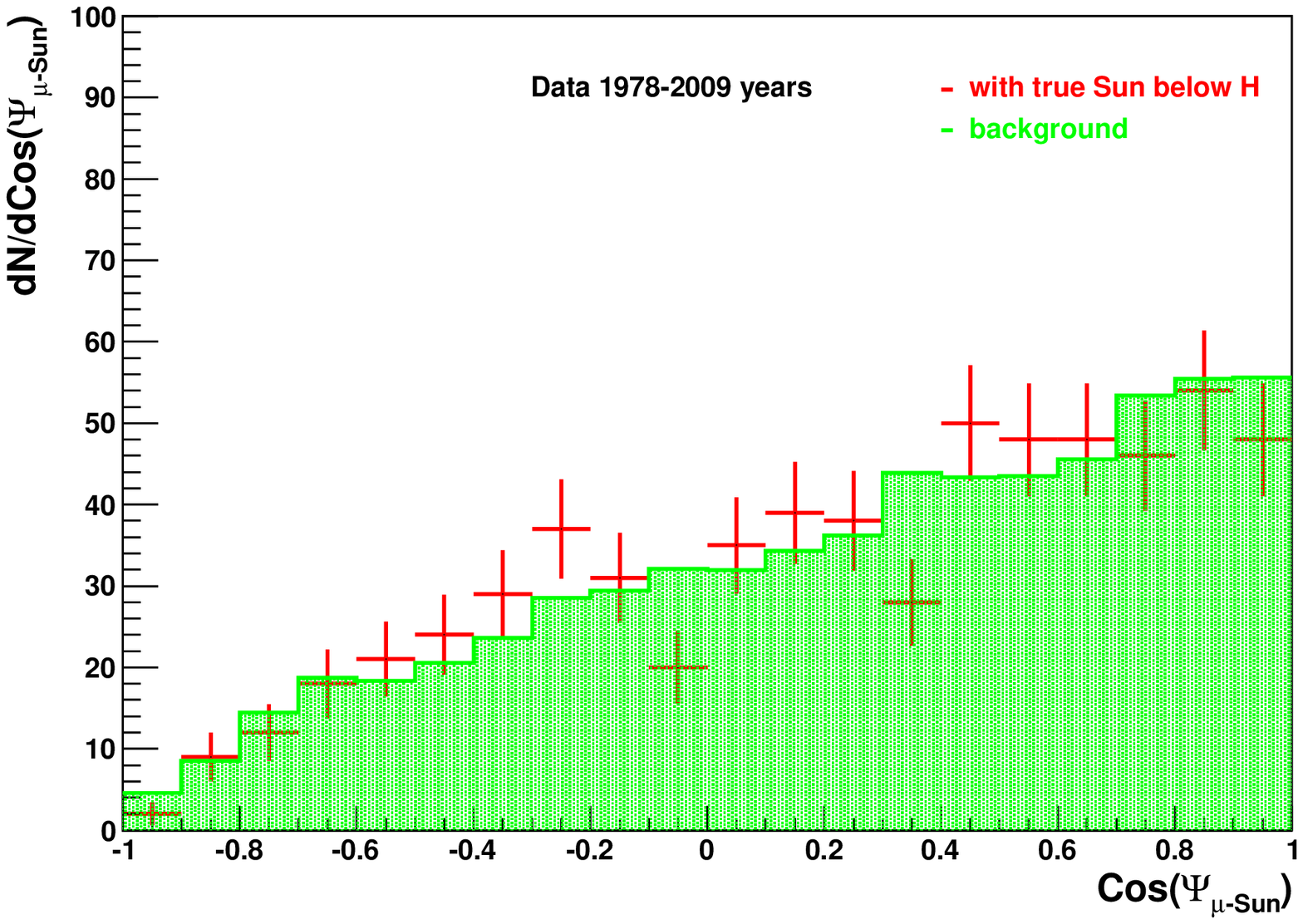} 
&
\includegraphics[angle=0,width=0.47\columnwidth]{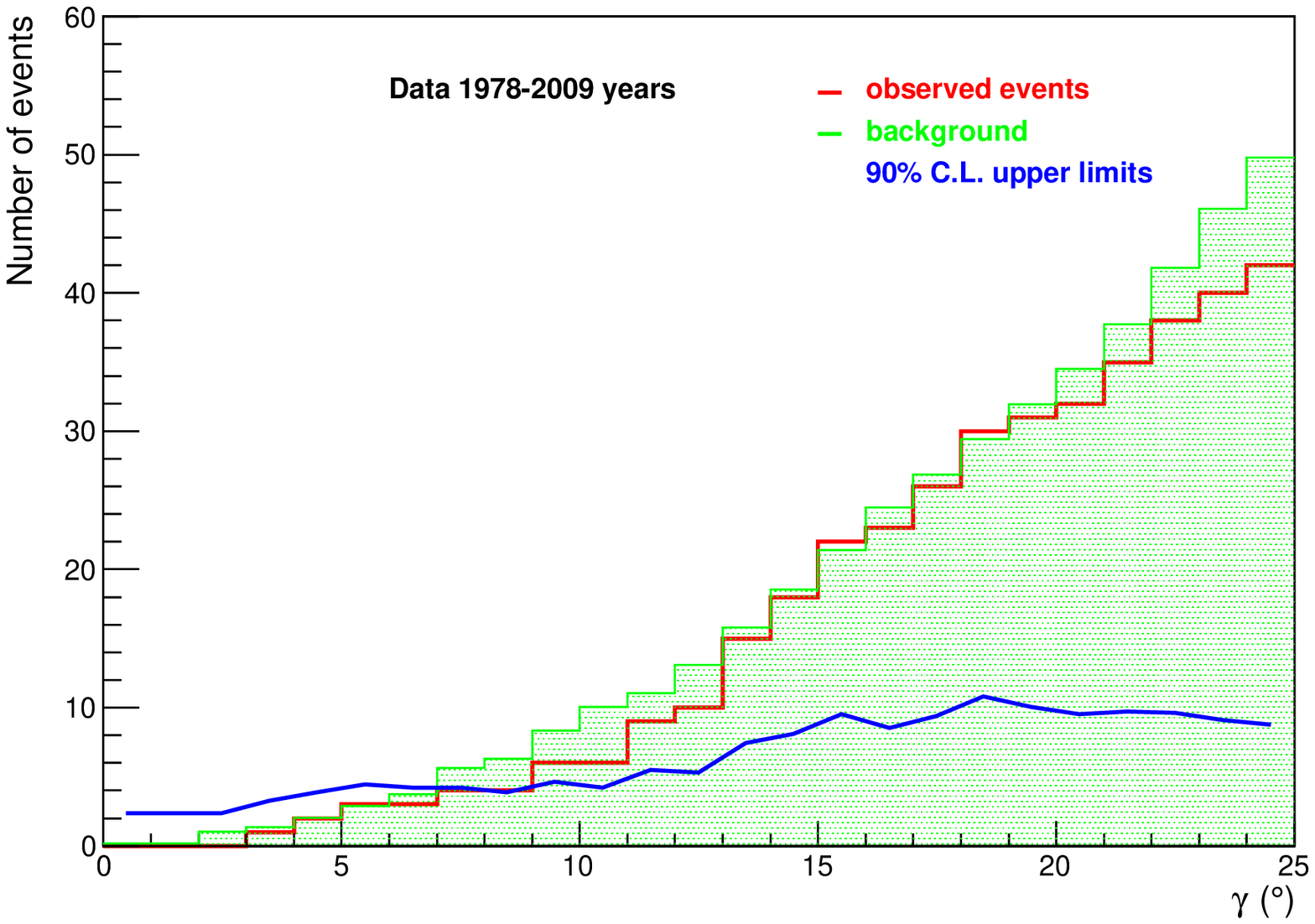} 
\end{tabular}
\caption{\label{AngLimToSun30-tot2009} Left: 
Distributions of $\cos{\Psi_{\mu-{\rm Sun}}}$ for
  events toward true Sun position below horizon (red) in comparison
  with measured background (green) obtained from the distributions for
  $\cos{\Psi_{\mu-{\rm Sun}}}$ averaged over six shifted Sun
  positions.
Right: The 90\% C.L. upper limits on additional number
of upward going muon events in the direction toward the Sun (blue) in
cone half-angle $\gamma$. Measured (red) and background (green) events
are also shown depending on the value of~$\gamma$.}  
}
\FIGURE[htb]{
\begin{tabular}{cc}
\includegraphics[angle=-90,width=0.45\columnwidth]{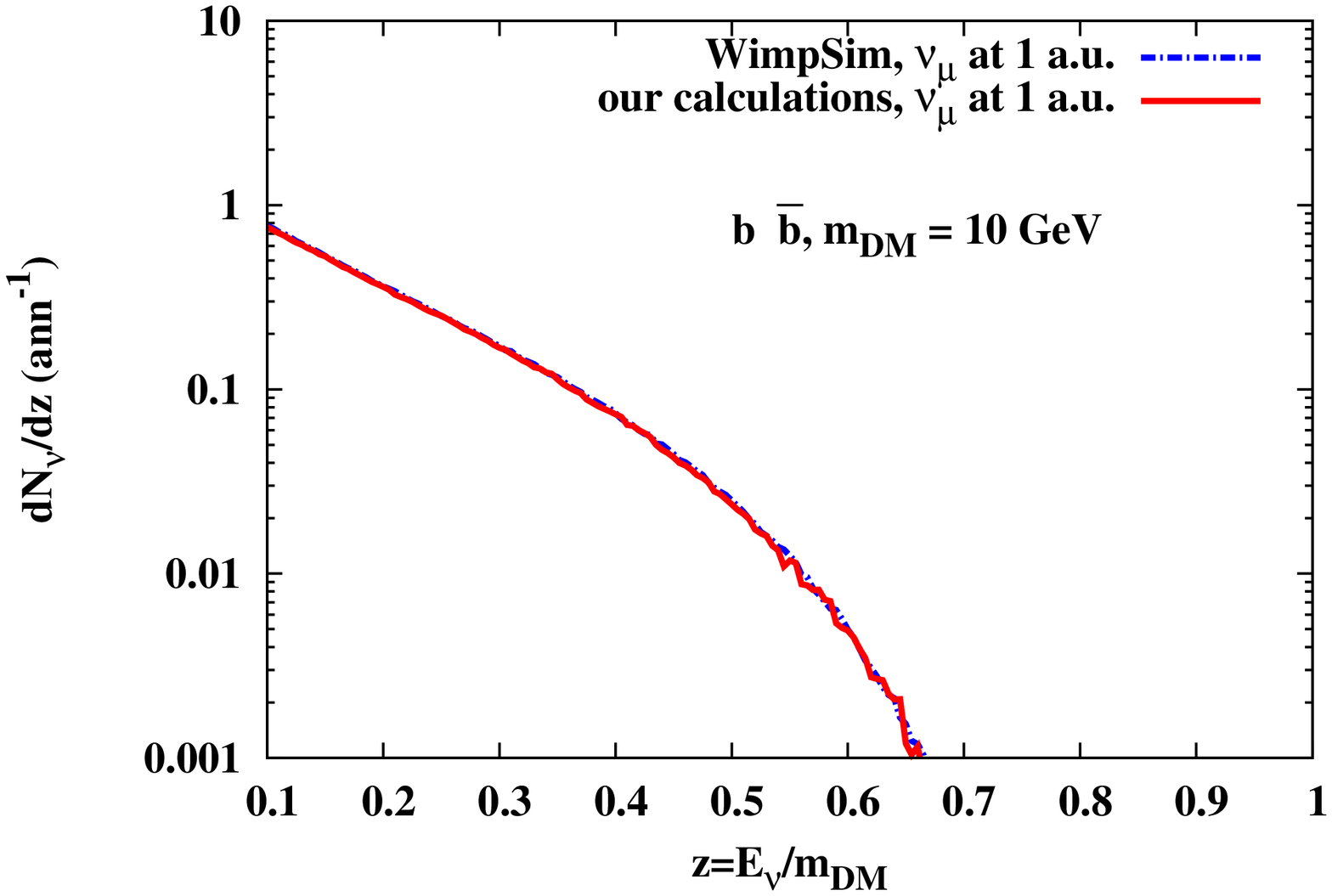} 
&
\includegraphics[angle=-90,width=0.45\columnwidth]{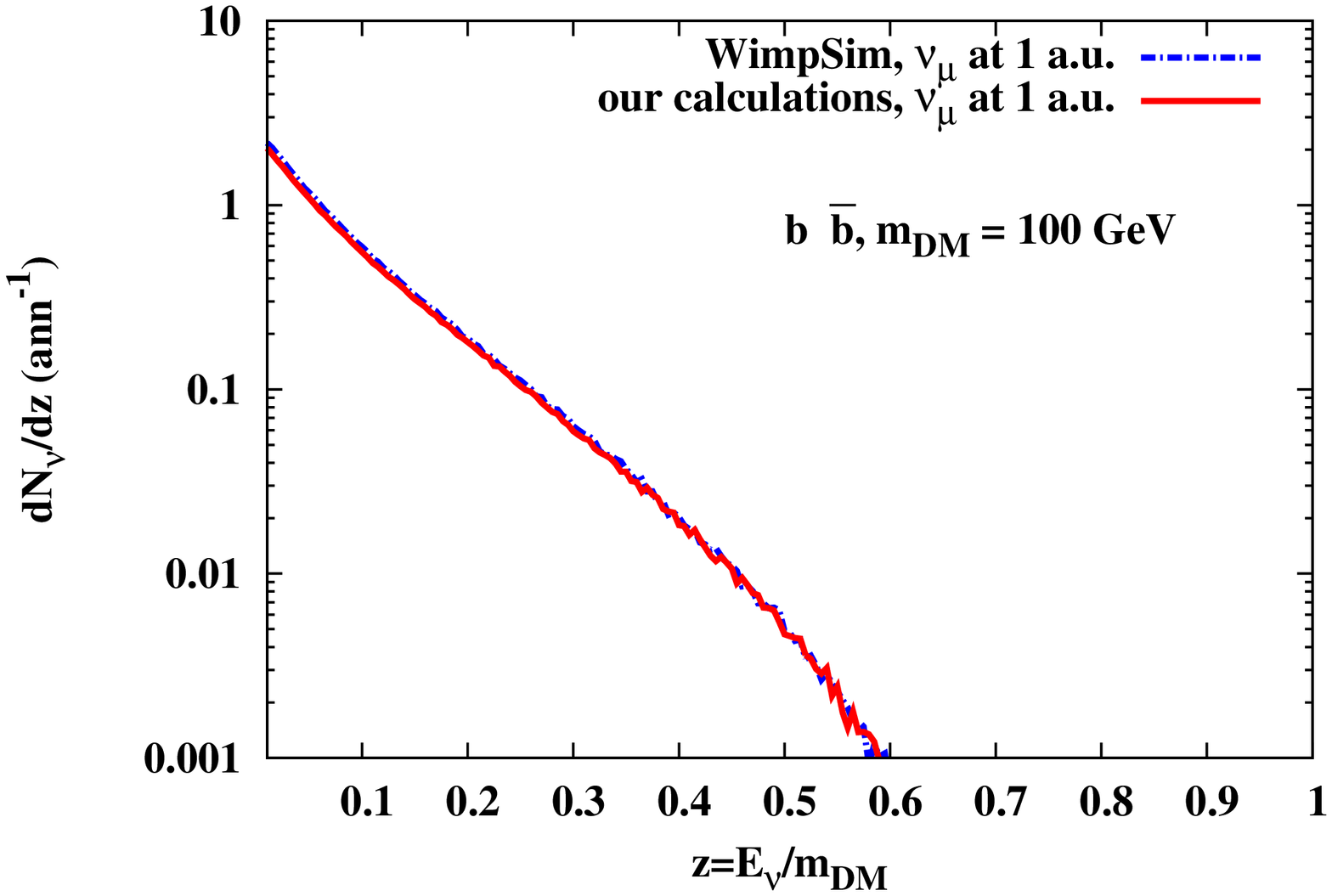} 
\end{tabular}
\caption{\label{wimpsim_comparison1} The energy spectra of $\nu_{\mu}$
  neutrino at production point and at the distance $R=$1 a.u. from the
  Sun obtained with the help of WimpSim package and by our
  calculations for $b\bar{b}$ annihilation channels and 
  $m_{\rm DM} = 10$~GeV (left) and $100$~GeV (right). 
}
}
\FIGURE[htb]{
\begin{tabular}{cc}
\includegraphics[angle=-90,width=0.45\columnwidth]{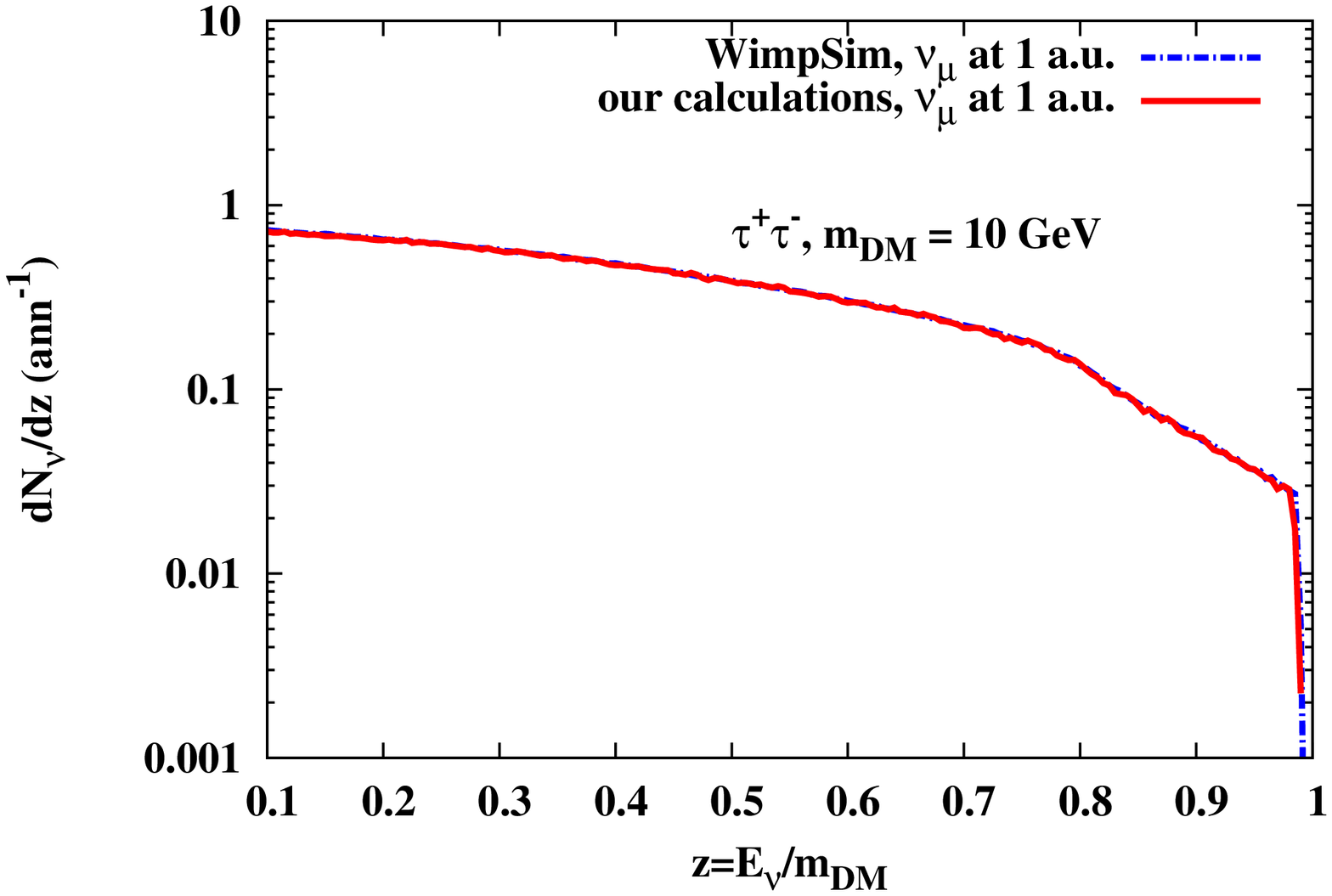} 
&
\includegraphics[angle=-90,width=0.45\columnwidth]{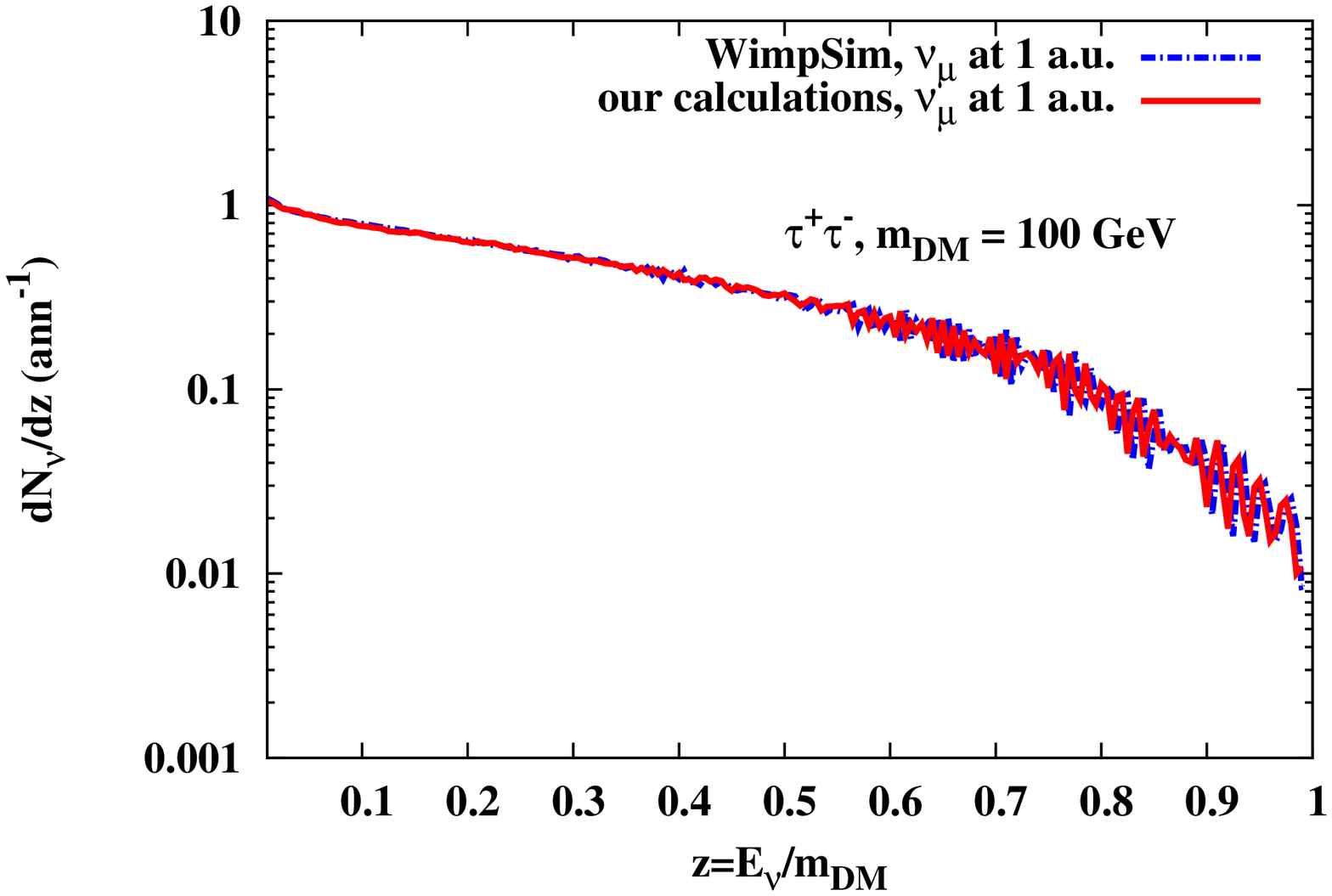} 
\end{tabular}
\caption{\label{wimpsim_comparison2} The energy spectra of $\nu_\tau$
  neutrino at production point and $\nu_{\mu}$ neutrino at the
  distance $R=1$~a.u. from the Sun obtained with the help of WimpSim
  package and by our calculations for $\tau^{+}\tau^{-}$ annihilation
  channels and $m_{\rm DM} = 10$~GeV (left) and $100$~GeV (right). 
}
}
\FIGURE[htb]{
\begin{tabular}{cc}
\includegraphics[angle=-90,width=0.45\columnwidth]{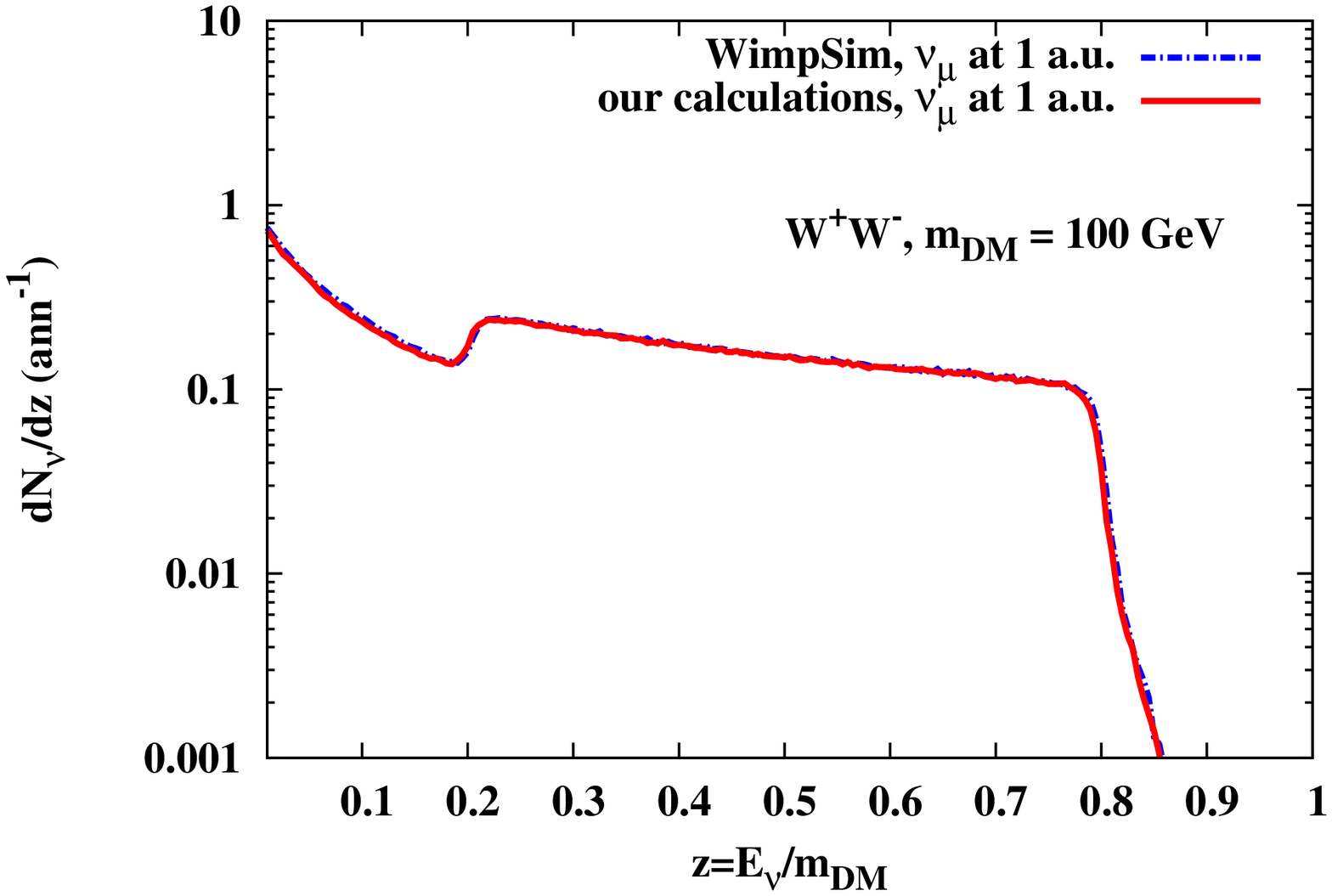} 
&
\includegraphics[angle=-90,width=0.45\columnwidth]{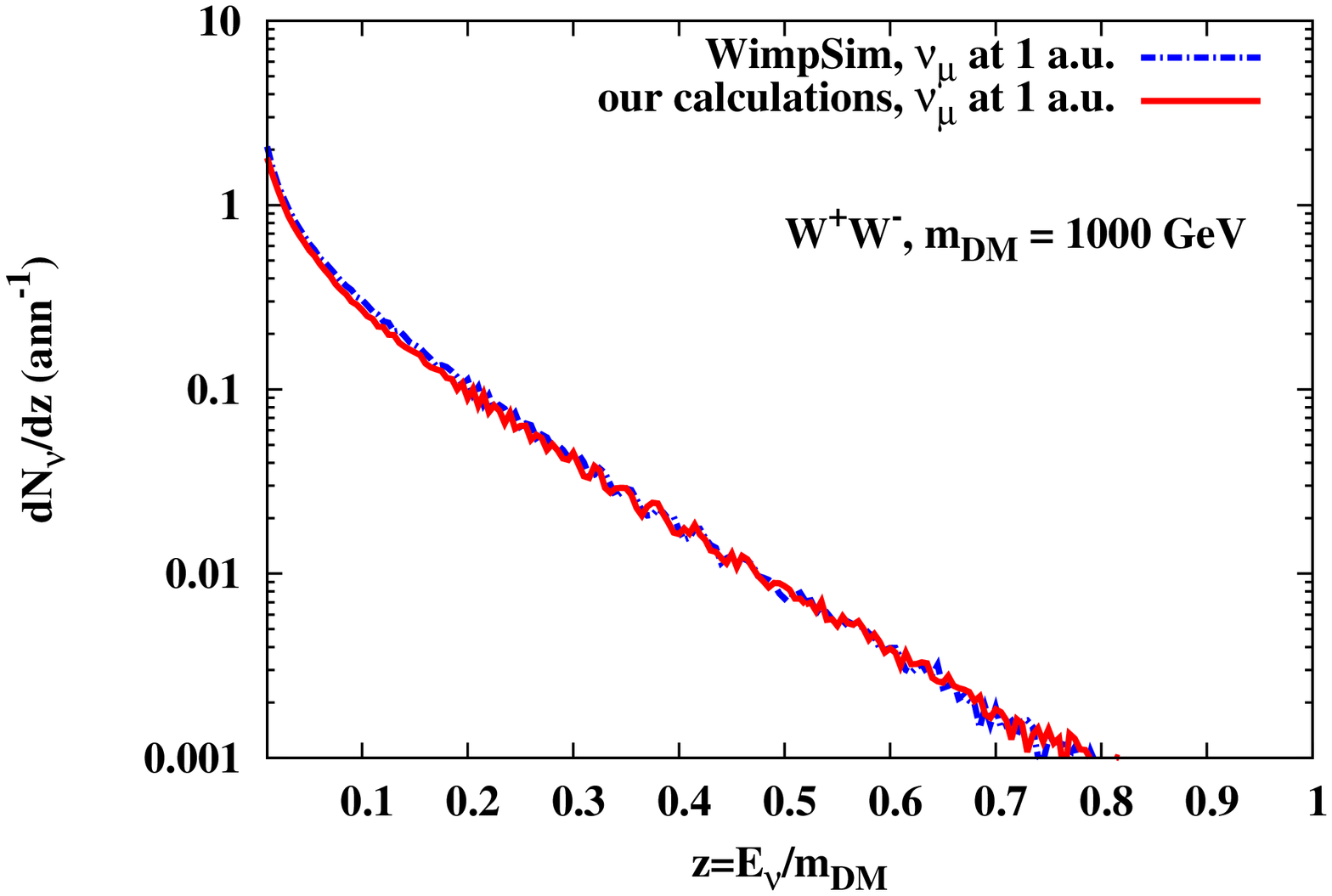} 
\end{tabular}
\caption{\label{wimpsim_comparison3} The energy spectra of $\nu_\tau$
  neutrino at production point and $\nu_{\mu}$ neutrino at the
  distance $R=1$~a.u. from the Sun obtained with the help of WimpSim
  package and by our calculations for $W^{+}W^{-}$ annihilation
  channels and $m_{\rm DM} = 100$~GeV (left) and $1000$~GeV (right).
}
}
\FIGURE[htb]{
\begin{tabular}{cc}
\includegraphics[angle=-90,width=0.45\columnwidth]{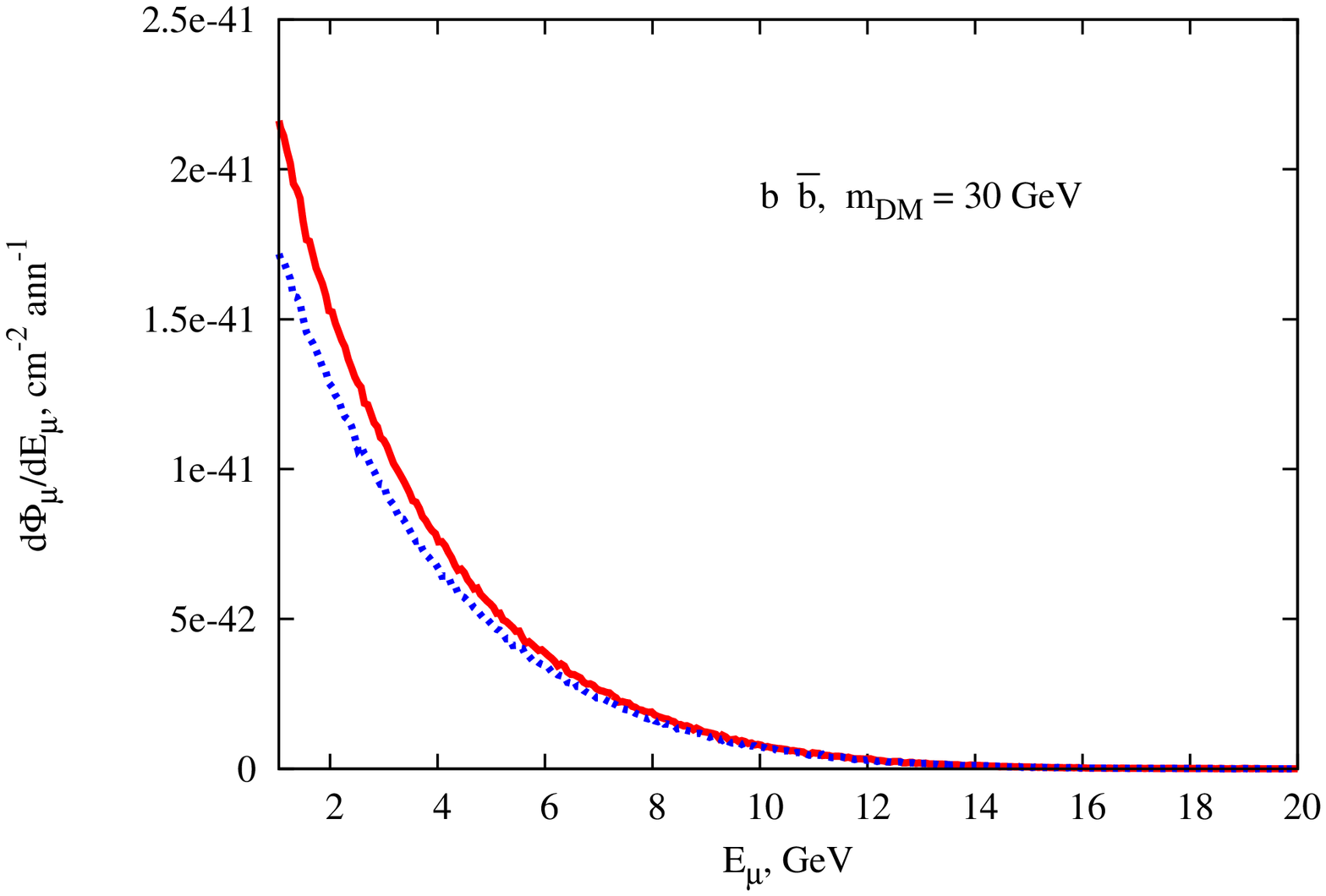} 
&
\includegraphics[angle=-90,width=0.45\columnwidth]{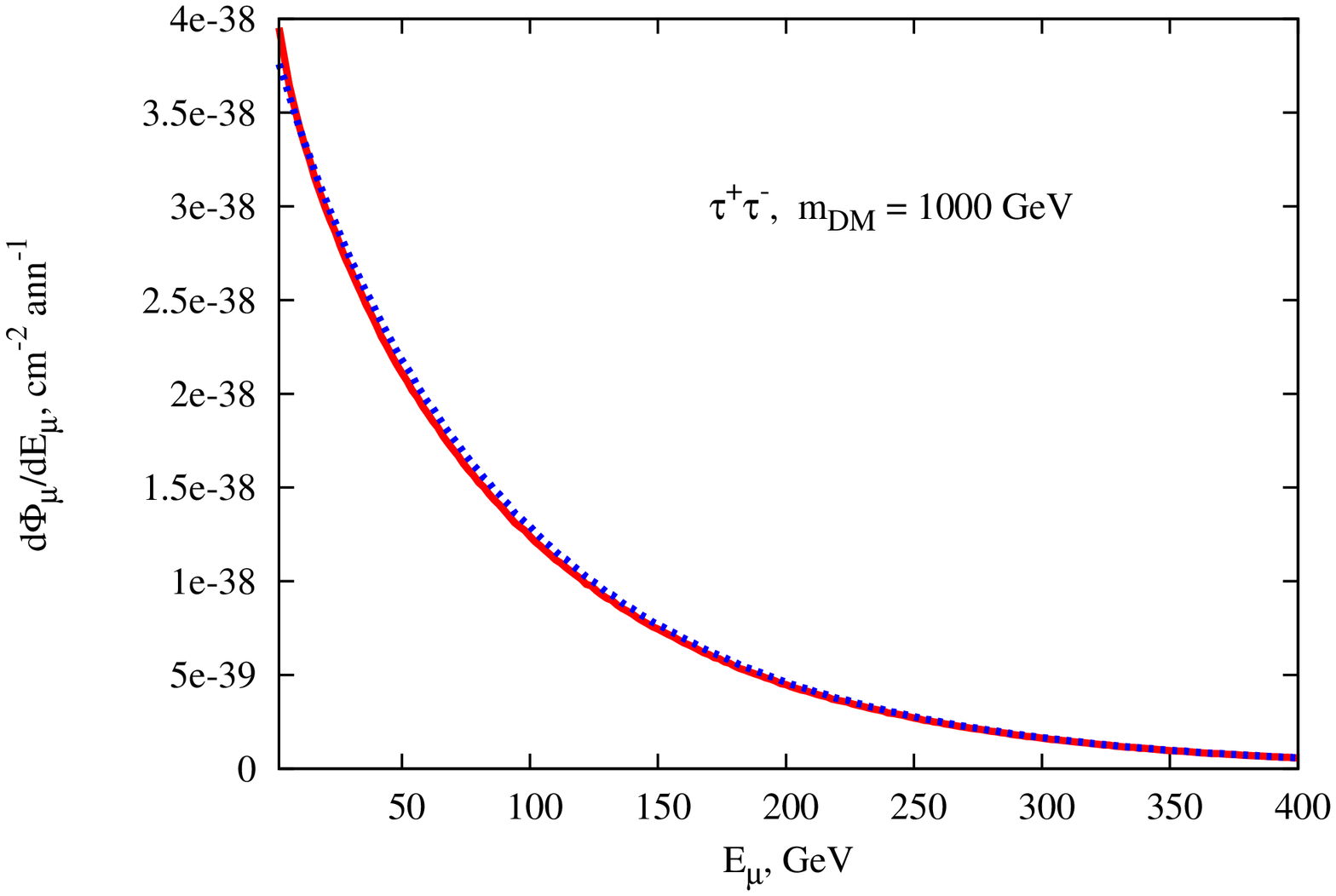} 
\end{tabular}
\caption{\label{muon_losses} The comparison between muon energy
  spectra obtained taking into account energy dependence for
  $\alpha(E)$ and $\beta(E)$ (red, solid line) and using constant
  values $\alpha(E)\approx 2.2\cdot 10^{-3}$~GeV$\cdot$cm$^2/$g  and
  $\beta(E)\approx 4.4\cdot 10^{-6}$~cm$^2/$g (blue, dotted line). 
}
}
\FIGURE[htb]{
\begin{tabular}{cc}
\includegraphics[angle=-90,width=0.45\columnwidth]{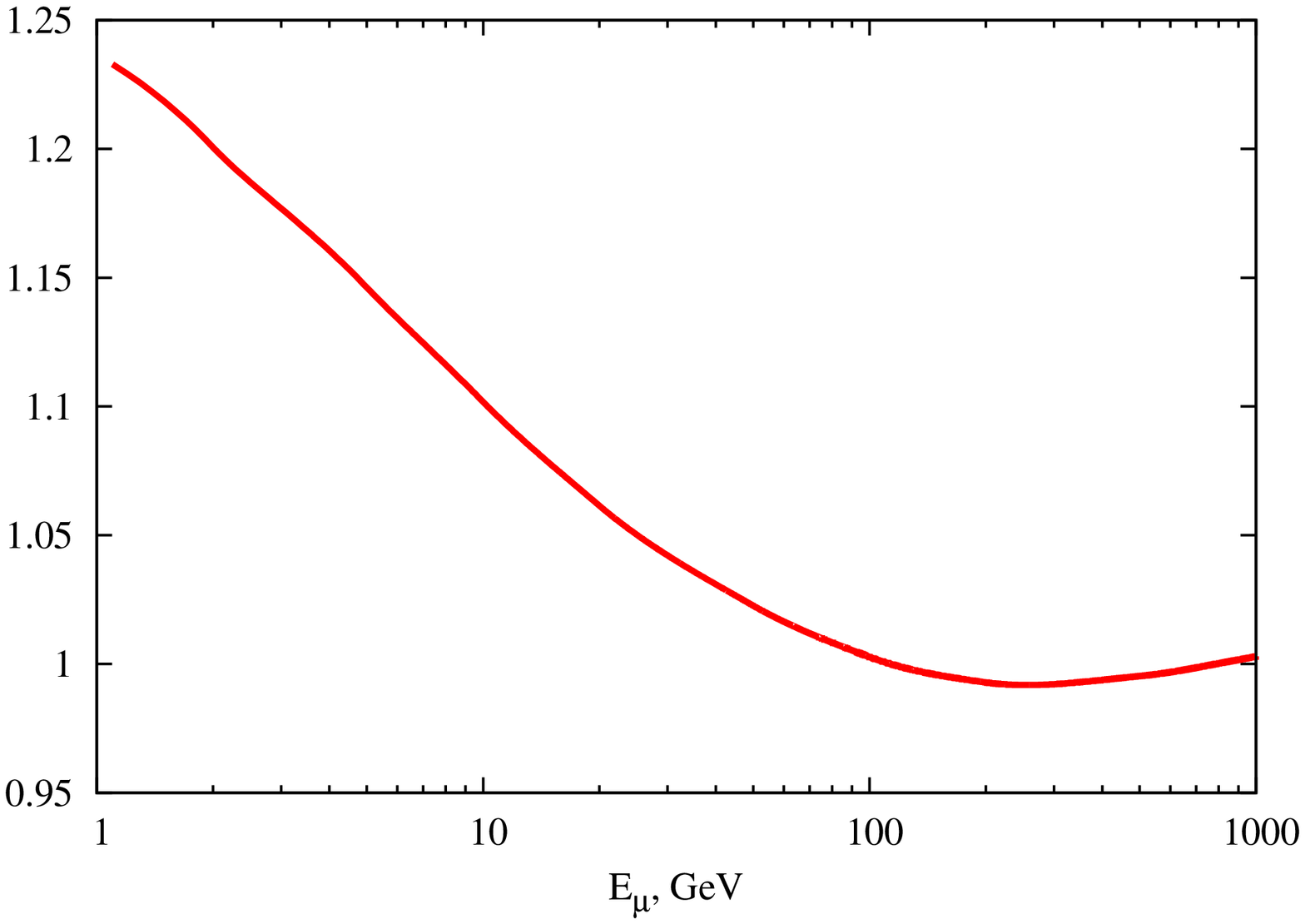} 
&
\includegraphics[angle=-90,width=0.45\columnwidth]{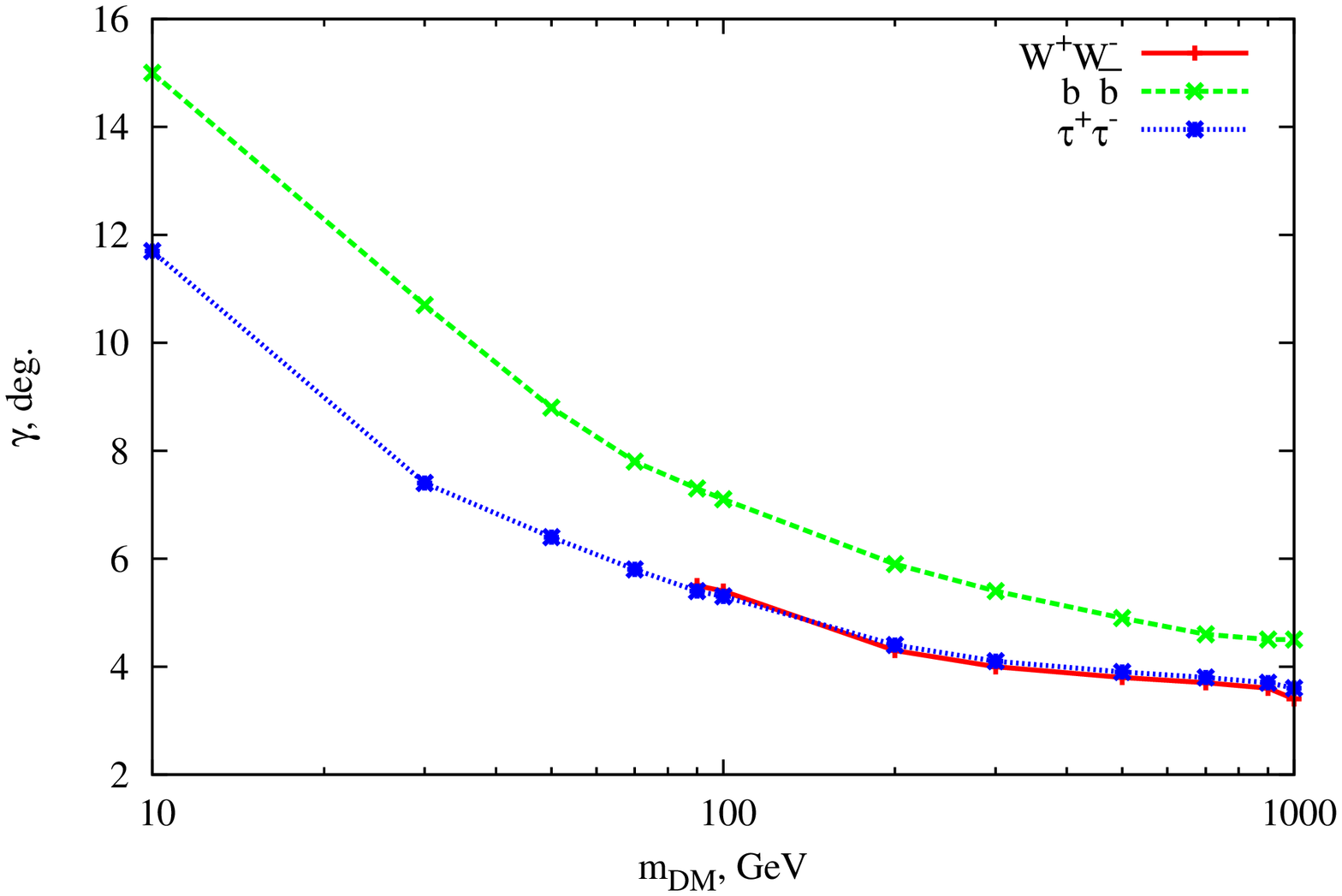}
\end{tabular}
\caption{\label{muon_ranges} Left: The ratio of muon ranges calculated
  using full energy dependence for $\alpha(E)$ and $\beta(E)$ and
  using constant values $\alpha(E)\approx 2.2\cdot
  10^{-3}$~GeV$\cdot$cm$^2/$g  and $\beta(E)\approx 4.4\cdot
  10^{-6}$~cm$^2/$g. Right: Cone half-angle $\gamma$ as a function of 
  the mass $m_{\rm DM}$ for three annihilations channels $b\bar{b}$,
  $\tau^{+}\tau^{-}$ and $W^{+}W^{-}$. 
}
}
\FIGURE[htb]{
\begin{tabular}{cc}
\includegraphics[angle=-90,width=0.45\columnwidth]{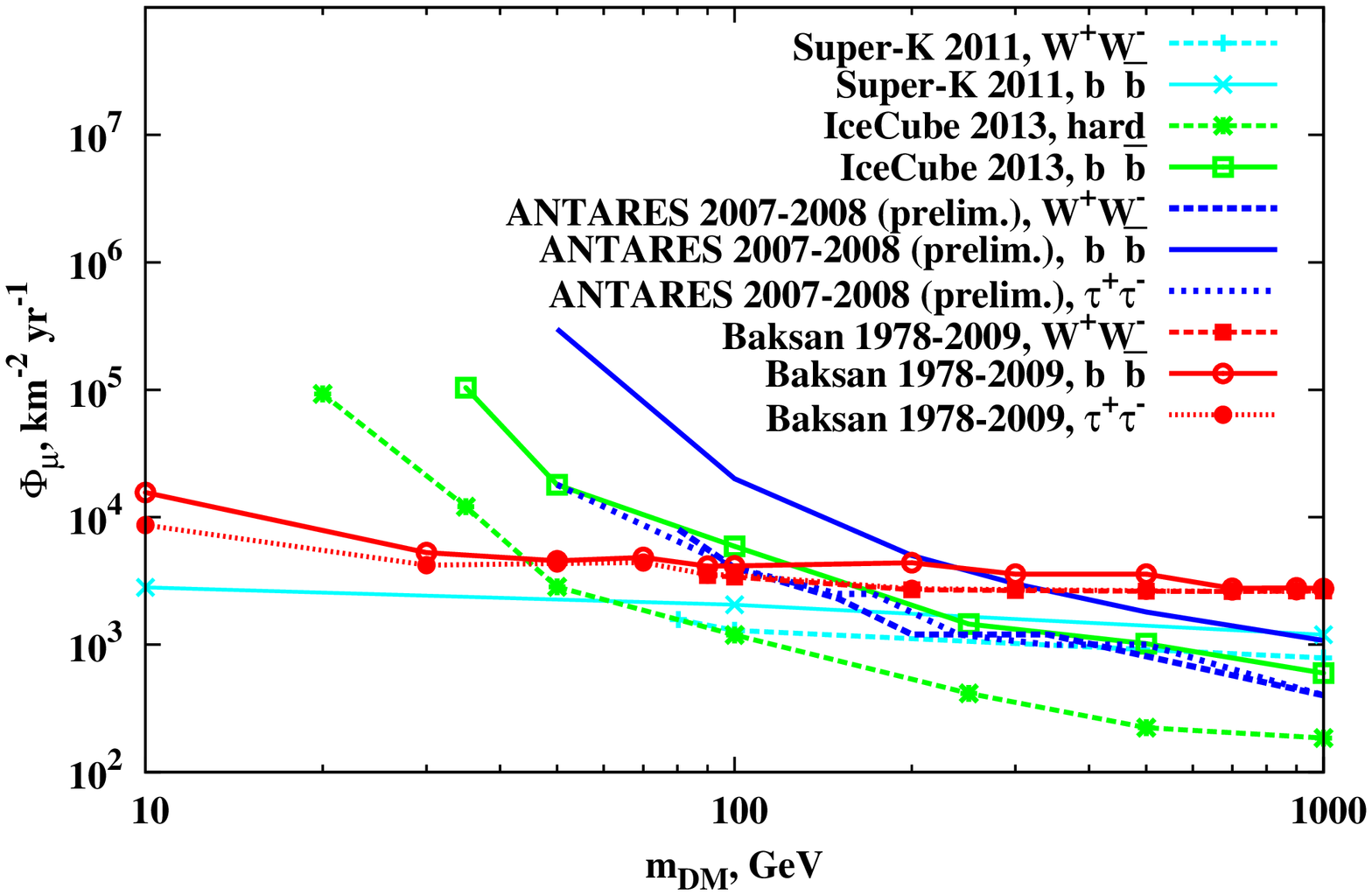} 
&
\includegraphics[angle=-90,width=0.45\columnwidth]{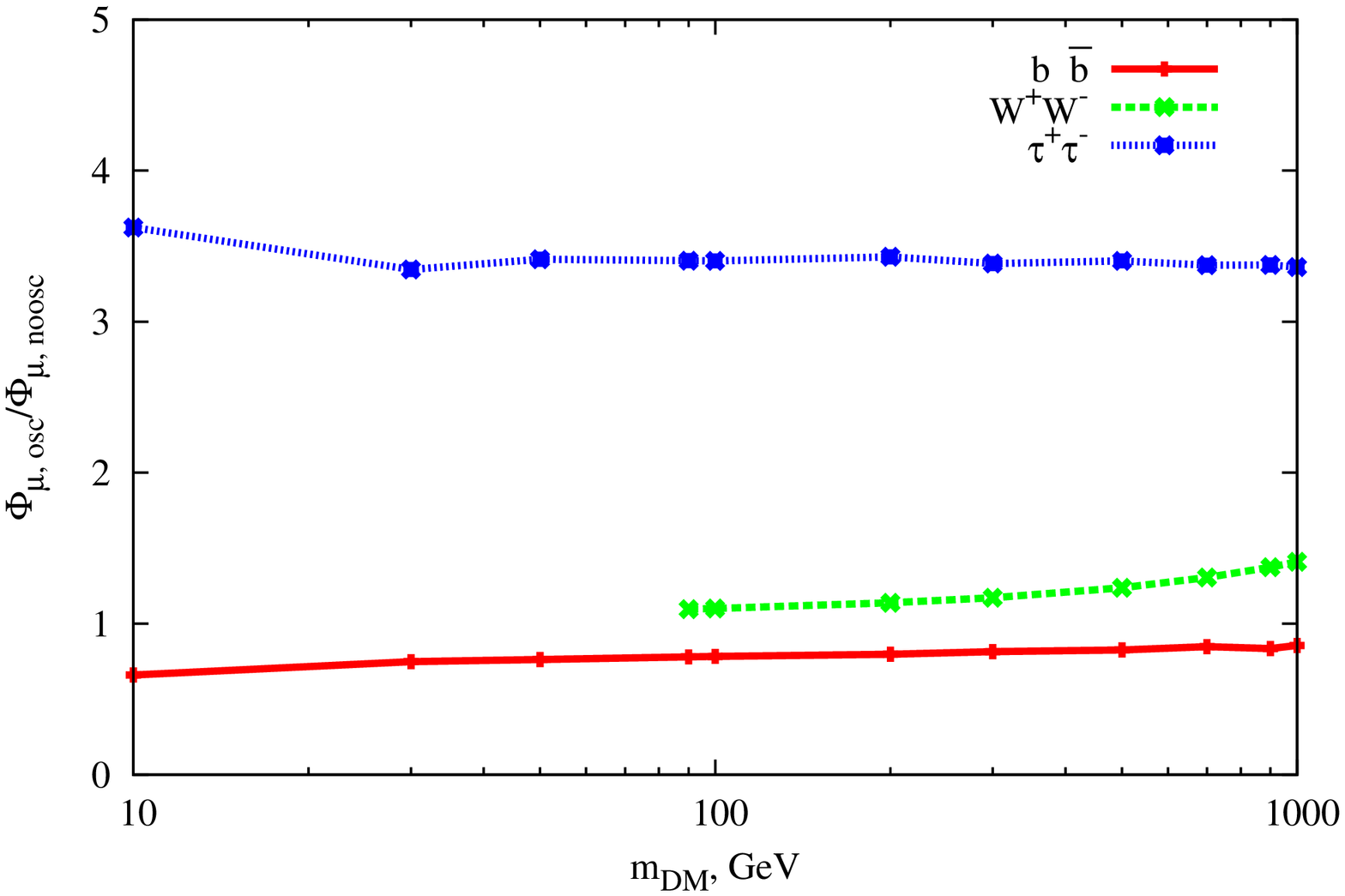}
\end{tabular}
\caption{\label{osc} The Baksan limits on muon flux from the dark
  matter   annihilations in the Sun in comparison with other
  experimental   results (left);  ratios of expected muon fluxes both
  from DM 
  neutrino and antineutrino generated in pure three annihilation
  branches $b\bar{b}$ quarks, or ${\tau^+\tau^-}$ leptons or
  ${W^+W^-}$ bosons in cases with and without three flavours
  oscillations (right).  
}
}
\FIGURE[htb]{
\includegraphics[angle=-90,width=1.0\columnwidth]{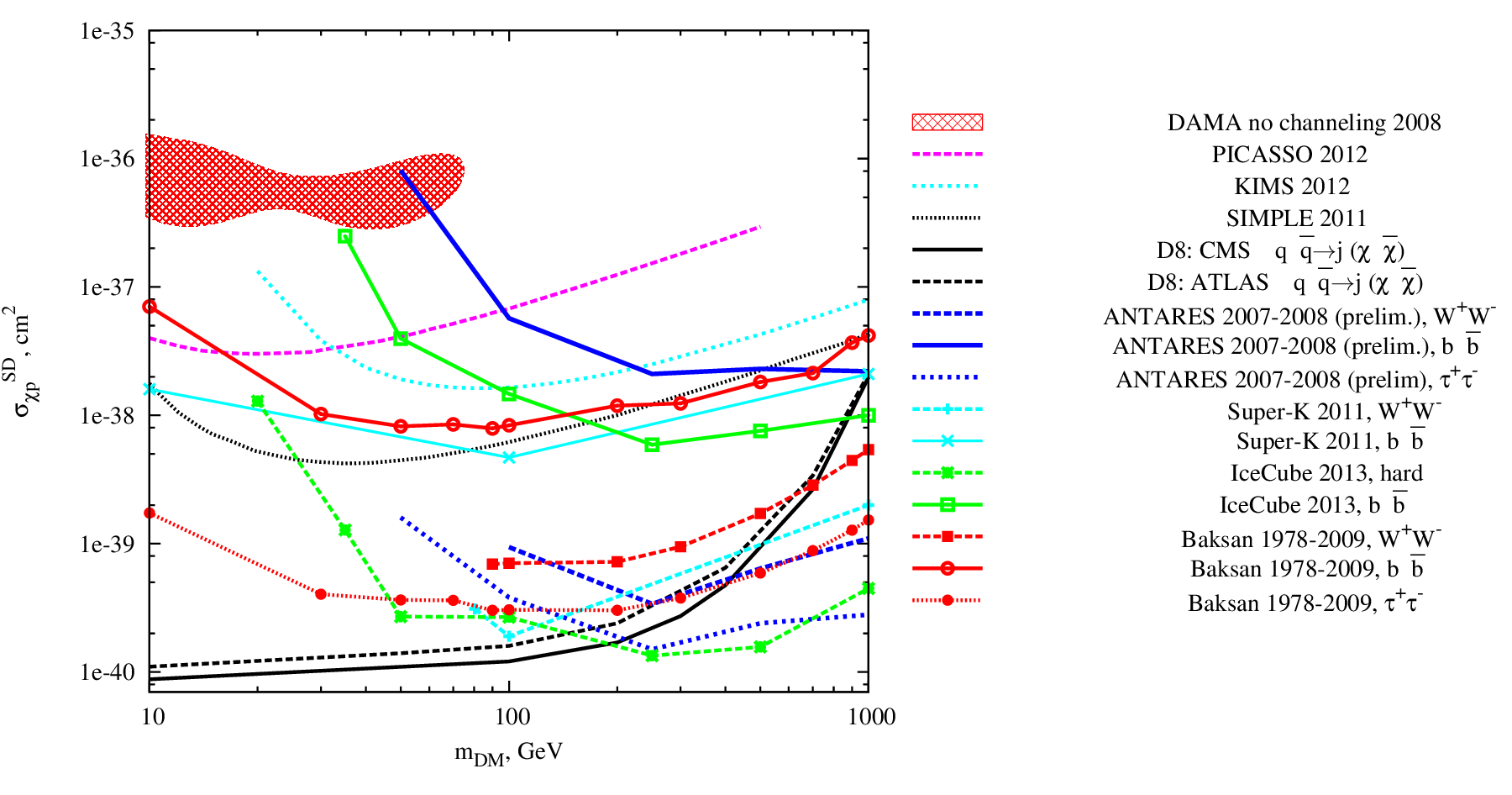} 
\caption{\label{sd_lim} The Baksan limits on SD elastic cross section
  of dark matter particle on proton in comparison with other
  experimental results.  We implement the DMTools~\cite{DMTools} 
database to plot the results of direct searches.
}
}
\FIGURE[htb]{
\includegraphics[angle=-90,width=1.0\columnwidth]{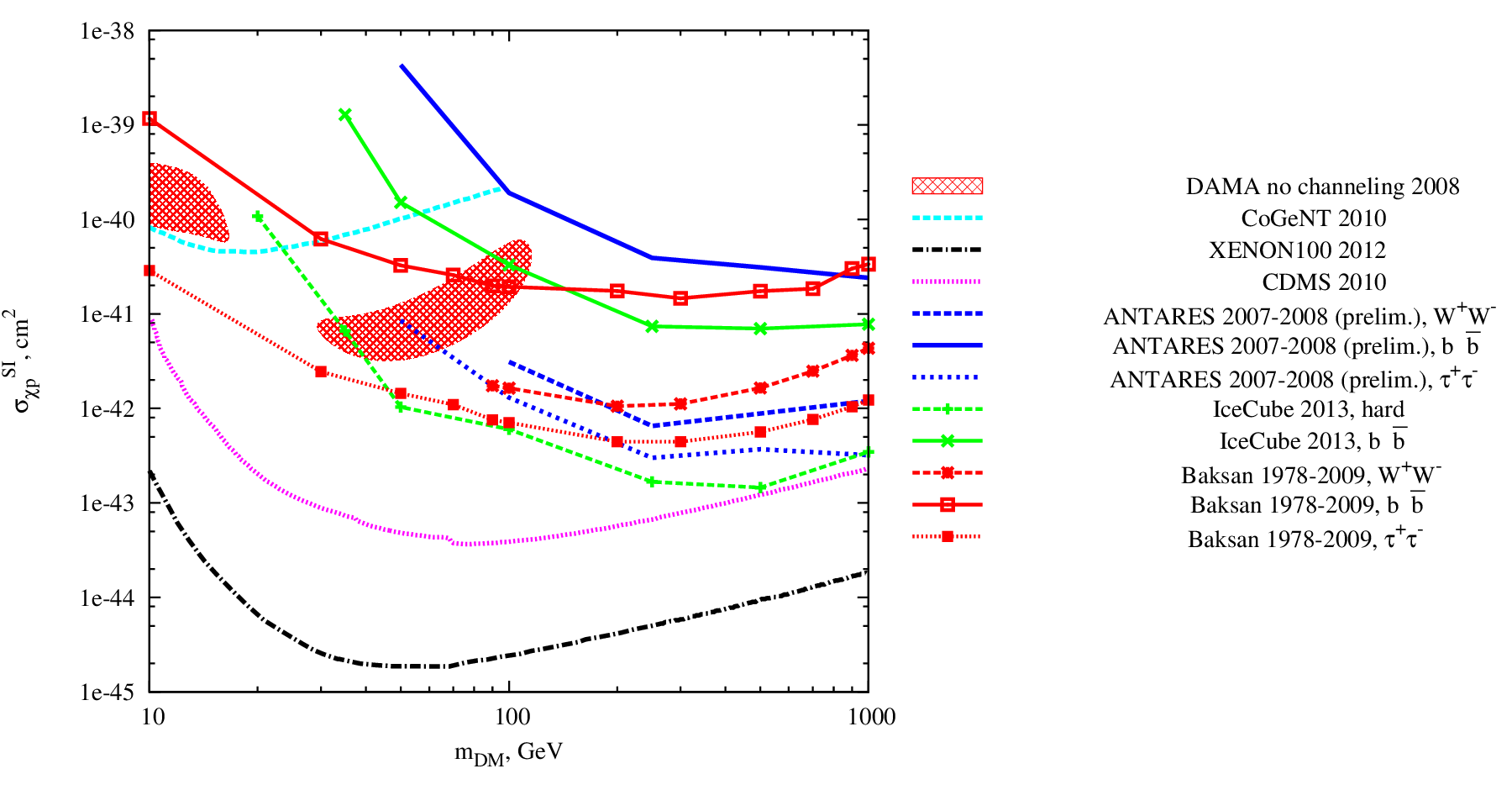} 
\caption{\label{si_lim} The Baksan limits on SI elastic cross section
  of dark matter particle on proton in comparison with other
  experimental results.  We implement the DMTools~\cite{DMTools} 
database to plot the results of direct searches.
}
}
\end{document}